\documentclass{aa}
\usepackage{txfonts}
\usepackage{natbib}
\usepackage{graphicx}
\usepackage{graphicx}
\usepackage{subcaption}
\usepackage{orcidlink}
\usepackage{hyperref}
\usepackage{listings}
\usepackage{xcolor}
\usepackage{upgreek}

\title{Mapping the Milky Way with Gaia Bp/Rp spectra \uppercase\expandafter{\romannumeral1}: Systematic flux corrections and atmospheric parameters for $68$ million stars}
\authorrunning{Ye et al.}
\titlerunning{Bp/Rp spectral corrections and atmospheric parameters for $68$ million stars}

\author{
    Xianhao Ye\orcidlink{0000-0002-5805-8112}\inst{\ref{inst:naoc},\ref{inst:iac},\ref{inst:ull}},
    Wenbo Wu\orcidlink{0000-0002-3354-9492}\fnmsep\thanks{The second author also made substantial contributions to the paper.}\inst{\ref{inst:naoc},\ref{inst:iac},\ref{inst:ull}},
    Carlos Allende Prieto\orcidlink{0000-0002-0084-572X}\fnmsep\thanks{Corresponding author, \email{carlos.allende.prieto@iac.es}}\inst{\ref{inst:iac},\ref{inst:ull}},
    David S. Aguado\orcidlink{0000-0001-5200-3973}\inst{\ref{inst:iac},\ref{inst:ull}},
    Jingkun Zhao\orcidlink{0000-0003-2868-8276}\inst{\ref{inst:naoc}},
    Jonay I. Gonz\'{a}lez Hern\'{a}ndez\orcidlink{0000-0002-0264-7356}\inst{\ref{inst:iac},\ref{inst:ull}},
    Rafael Rebolo\orcidlink{0000-0003-3767-7085}\inst{\ref{inst:iac},\ref{inst:ull}},
    Gang Zhao\orcidlink{0000-0002-8980-945X}\fnmsep\thanks{Corresponding author, \email{gzhao@nao.cas.cn}}\inst{\ref{inst:naoc},\ref{inst:ucas}},
    Zhuohan Li\orcidlink{0000-0002-1126-9289}\inst{\ref{inst:naoc},\ref{inst:ucas}},
    Carlos del Burgo\orcidlink{0000-0002-8949-5200}\inst{\ref{inst:iac},\ref{inst:ull}},
    Yuqin Chen\orcidlink{0000-0002-8442-901X}\inst{\ref{inst:naoc}}
}

\institute{
    National Astronomical Observatories, Chinese Academy of Sciences, Beijing 100101, People's Republic of China\label{inst:naoc}
    \and
    Instituto de Astrof\'{i}sica de Canarias, V\'{i}a L\'{a}ctea, 38205 La Laguna, Tenerife, Spain\label{inst:iac}
    \and
    Universidad de La Laguna, Departamento de Astrof\'{i}sica, 38206 La Laguna, Tenerife, Spain\label{inst:ull}
    \and
    School of Astronomy and Space Science, University of Chinese Academy of Sciences, Beijing 100049, People's Republic of China\label{inst:ucas}
}

\date{Received November 14, 2024; accepted Janurary 13, 2025}

\abstract
  {Gaia Bp/Rp spectrophotometry for over two hundred million stars has been publicly released as part of Gaia Data Release 3 (DR3). These data have great potential for mapping metallicity across the Milky Way. Several recent studies have analyzed this data set to derive atmospheric parameters and identify new metal-poor stars. In addition, systematics in the fluxes of the Bp/Rp spectra have also been identified and characterized.}
  {We aim to construct an alternative catalog of atmospheric parameters from Gaia Bp/Rp spectra by fitting them with synthetic spectra based on model atmospheres, and provide corrections to the Bp/Rp fluxes according to stellar colors, magnitudes, and interstellar extinction.}
  {We use \texttt{GaiaXPy} to obtain calibrated spectra and apply \texttt{FER\reflectbox{R}}E to match the corrected Bp/Rp spectra with models and infer atmospheric parameters. We train a neural network (NN) using stars in the Apache Point Observatory Galactic Evolution Experiment (APOGEE) to predict flux corrections as a function of wavelength for each target.}
  {Based on the comparison with APOGEE parameters, we conclude that our estimated parameters have systematic errors and uncertainties in $T_{\mathrm{eff}}$, $\log g$, and [M/H] about $-38 \pm 167$ K, $0.05 \pm 0.40$ dex, and $-0.12 \pm 0.19$ dex, respectively, for stars in the range $4000 \le T_{\mathrm{eff}} \le 7000$ K. The corrected Bp/Rp spectra show improved agreement with both models and Hubble Space Telescope (HST) CALSPEC data. Our correction increases the precision of the relative spectrophotometry of the Bp/Rp data from $3.2\% - 3.7\%$ to $1.2\% - 2.4\%$. We also compare our results with other similar catalogs from the literature and validate them using star clusters. Finally, we have built a catalog of atmospheric parameters for stars within $4000 \le T_{\mathrm{eff}} \le 7000$ K, comprising $68,394,431$ sources, along with a subset of $124,188$ stars with $\mathrm{[M/H]} \le -2.5$. Our catalogs and flux correction code are publicly available.}
  {Our results confirm that the Gaia Bp/Rp flux calibrated spectra show systematic patterns as a function of wavelength that are tightly related to colors, magnitudes, and extinction. Our optimization algorithm can give us accurate atmospheric parameters of stars with a clear and direct link to models of stellar atmospheres, and can be used to efficiently search for extremely metal-poor (EMP) stars.}

\keywords{Techniques: spectroscopic-Catalogs-Stars: fundamental parameters}

\begin{document}

\maketitle

\section{Introduction}\label{sec:intro}

The stellar metallicities [M/H] keep track of the formation and evolution of the Milky Way. Due to the chemical enrichment of the interstellar medium by successive stellar generations, [M/H] can be used as a proxy of age and reflects the birth environment of a star. Assuming that metal-rich disk stars are born on nearly circular orbits in the Galactic plane from chemically well-mixed cold gas, mono-abundance stellar populations with the same metal content ([M/H], $[\alpha/\text{Fe}]$) shall be formed at the same birth radius $R_\text{b}$ and look-back time $\tau$ \citep{2009MNRAS.399.1145S}. This assumption has been used to establish a comprehensive chemodynamical model of our Galaxy \citep{2021MNRAS.507.5882S,2022MNRAS.513.4130L,2023ApJ...954..124I,2023MNRAS.523.3791C,2024MNRAS.527.1915B}. Combined with kinematics, the metallicities of metal-poor halo stars serve as a powerful tool to separate in-situ populations from the remnants of ancient accretion events \citep{2020ApJ...901...48N,2022MNRAS.514..689B,2022arXiv220402989C,2024NewAR..9901706D}. These newly found accreted substructures are useful to check the $\Lambda$CDM cosmological paradigm in which Milky Way–sized halos are built from the mergers of smaller satellite galaxies. Extremely metal-poor (EMP) stars may have formed from gas enriched only by the very first stars, also called Pop \uppercase\expandafter{\romannumeral3} \citep{2023ARA&A..61...65K}. Therefore, their abundance patterns hold the key to constrain nucleosynthesis in the earliest supernova events \citep{2006NuPhA.777..424N,2010ApJ...724..341H,2014Sci...345..912A,2013ARA&A..51..457N,2023MNRAS.525..190K}. The low-metallicity tail of the metallicity distribution function (MDF) provides essential constrains on the chemical enrichment in the early phases of the formation of the Milky Way \citep{2007MNRAS.381..647S,2010ApJ...717..542K,2013MNRAS.436.1362Y,2019ApJ...871..206S,2020ApJ...897...58T,2020MNRAS.492.4986Y}. However, at present only a few hundred stars with $\mathrm{[M/H]}<-3$ and significantly fewer with $\mathrm{[M/H]}<-4$ have been identified, and these are insufficient to give a complete picture.

We are witnessing an era in which stellar atmospheric parameters are available for massive samples thanks to large spectroscopic surveys, such as the Sloan Digital Sky Survey (SDSS) \uppercase\expandafter{\romannumeral1}-\uppercase\expandafter{\romannumeral4} \citep{2000AJ....120.1579Y,2004AJ....128..502A,2011ApJS..193...29A,2022ApJS..259...35A}, the Large Sky Area Multi-Object Fiber Spectroscopic Telescope (LAMOST) survey \citep{2006ChJAA...6..265Z,2012RAA....12..723Z,2012RAA....12.1197C,2012RAA....12.1243L}, the GALactic Archaeology with HERMES\footnote[1]{High Efficiency and Resolution Multi Element Spectrograph} (GALAH) survey \citep{2015MNRAS.449.2604D,2021MNRAS.506..150B}, the SDSS \uppercase\expandafter{\romannumeral5} \citep{2017arXiv171103234K}, the Dark Energy Spectroscopic Instrument (DESI) survey \citep{2023ApJ...947...37C}, the 4-metre Multi-Object Spectroscopic Telescope (4MOST) survey \citep{2019Msngr.175....3D}, or the WHT\footnote[2]{William Herschell Telescope} Enhanced Area Velocity Explorer (WEAVE) survey \citep{2024MNRAS.530.2688J}. These surveys have already accumulated tens of millions of stellar spectra, and continue to gather more. Combining the spectroscopic data with accurate measurements of proper motions and parallaxes provided by the European Space Agency (ESA) Gaia mission \citep{2016A&A...595A...2G,2023A&A...674A...1G}, enables us to carry out a detailed investigation of the chemodynamic properties of different stellar populations of the Milky Way.

In addition to the astrometric catalog, Gaia Data Release 3 \citep[DR3;][]{2023A&A...674A...1G} includes a catalog of 32.2 million high-resolution spectra ($R\sim8000$) centered on the near infrared Ca \uppercase\expandafter{\romannumeral2} triplet, obtained by the Radial Velocity Spectrometer (RVS) instrument \citep{2018A&A...616A...5C,2018A&A...616A...6S,2023A&A...674A..27A}. It also includes a catalog of 220 million low-resolution spectra, referred to as Bp/Rp spectra (hereafter XP spectra), obtained with the blue (wavelength range of 330-680 nm) and red (wavelength range of 640-1050 nm) Gaia slitless spectrophotometers, BP and RP, respectively \citep{2021A&A...652A..86C,2023A&A...674A...3M,2023A&A...674A...2D}. The XP spectra have 110 pixels and a variable resolving power, ranging from 20 to 90, as a function of wavelength. The XP spectra in DR3 are represented by Hermite basis functions \citep{2023A&A...674A...3M}.

The \texttt{Gaia Data Processing and Analysis Consortium} (DPAC) has developed the \texttt{Python} library \texttt{GaiaXPy}\footnote[3]{\href{https://gaiaxpy.readthedocs.io/en/latest/cite.html}{https://gaiaxpy.readthedocs.io/en/latest/cite.html}, DOI v2.1.0: 10.5281/zenodo.8239995} to facilitate handling XP spectra. This software allows the transformation of coefficients into calibrated spectra and photometry. Recent studies have shown that the synthetic photometry from XP spectra can be used to perform a highly precise calibration of other photometry surveys, such as Pristine \citep{2024A&A...692A.115M}, the Panoramic Survey Telescope and Rapid Response System (Pan-STARRS) \citep{2023ApJS..268...53X}, or the Javalambre-Photometric Local Universe Survey (J-PLUS) \citep{2023ApJS..269...58X,2024ApJ...972..112C}. However, other studies have revealed that XP calibrated spectra suffer from systematic errors that show up as "wiggles" \citep{2024ApJS..271...13H}, introduced by the combination of noise and the choice of data representation \citep{2023A&A...671A..52W}. The pattern of wiggles largely depends on the stellar color $G_{\mathrm{BP}}-G_{\mathrm{RP}}$ and the apparent magnitude $G$ \citep{2023A&A...674A...3M}, and may cause a modest bias in the synthetic photometry \cite{2024ApJS..271...13H} and the inferred stellar parameters. Due to its low-resolution and the aforementioned systematics, it is hard to determine chemical abundances from XP spectra. Nonetheless, we are still able to obtain a reliable estimation of the overall metallicity ([M/H]) down to the EMP domain, as shown using mock XP spectra in \cite{2022MNRAS.516.3254W}. Considering the large data volume and the full sky coverage, XP spectra provide us with a unique opportunity to build a complete metallicity map of the Milky Way.

Previous studies have adopted different methodologies to extract information from the low-resolution XP spectra. Traditional model-driven methods predict the atmospheric parameters by comparing the observed spectra with stellar spectral libraries \citep[e.g.,][]{1997A&A...318..841C,1998A&AS..130...65L,2005A&A...443..735C,2008A&A...486..951G,2013A&A...553A...6H,2018A&A...618A..25A}. As part of the astrophysical parameters inference system (Apsis), the Gaia General Stellar Parameterizer from Photometry (GSP-phot) adopts a Bayesian forward-modelling approach to fit XP spectra with isochrone models and four different stellar libraries \citep{2023A&A...674A..27A}. It provides an estimate of effective temperature $T_\mathrm{eff}$, surface gravity $\log g$, metallicity [M/H], absolute magnitude $M_{\mathrm{G}}$, radius, distance, and extinction for each star. However, the authors of GSP-Phot advise against using these metallicity estimates since they are dominated by large systematic errors.

\cite{2024ApJS..272...20A} constructed an all-sky 3D extinction map by comparing the low-resolution XP spectra with their empirically calibrated synthetic spectra and gave a reliable estimate of metallicity for stars with $\mathrm{[M/H]}>-1$. \cite{2024A&A...692A.115M} built a model linking the metallicity-sensitive synthetic CaHK magnitudes to the photometric metallicity $\mathrm{[M/H]_{phot}}$ from XP spectra. Even though only [M/H] measurements are provided, their results show good agreement with the literature in the full metallicity range, even for stars under $\mathrm{[M/H]}=-3$. \cite{2023A&A...674A.194B} and \cite{2024A&A...687A.177X} also tried to extract metallicity information from the synthetic photometry, but they only determined $\mathrm{[M/H]_{phot}}$ for a small fraction of the XP spectra available in DR3. The intrinsic nature of the estimation of $\mathrm{[M/H]_{phot}}$ also involves a comparison between synthetic spectra and observations, albeit indirectly, through the photometry generated from them. The accuracy of model-driven methods highly relies on the consistency between the XP spectra and the synthetic spectra. Unfortunately, there are several points that may induce discrepancies between them.

First, the theoretical spectra are limited by our knowledge of stellar atmospheres, atomic and molecular physics. Second, XP spectra are not perfectly flux-calibrated. Third, the information in the low-resolution spectra is highly sensitive to surface temperature $T_\mathrm{eff}$ and gravity $\log g$. The metallicity determined from XP spectra with low ($T_\mathrm{eff}<4000$ K) and high ($T_\mathrm{eff}>7000$ K) temperature are likely unreliable, due to the limited strength of metal lines in stars of such low metallicity and high effective temperature.

In the last decade, data-driven and machine learning methods have been widely adopted to overcome the gap between synthetic spectra and observations \citep[e.g.,][]{2015ApJ...808...16N,2017ApJ...849L...9T,2019ApJ...879...69T,2019ApJS..245...34X,2019MNRAS.483.3255L}. Data-driven methods use a large training sample with known labels to construct a relationship between the stellar parameters and the spectra. Previous studies find that data-driven methods are effective to derive precise atmospheric parameters and elemental abundances for low-resolution spectra by applying them to the LAMOST survey \citep{2019PASP..131e5001W,2022MNRAS.517.4875L,2023MNRAS.521.6354L}. Therefore, it is natural to turn to data-driven methods to infer stellar properties from XP spectra \citep[e.g.,][]{2022ApJ...941...45R,2023ApJS..267....8A,2023MNRAS.524.1855Z,2023MNRAS.521.2745S,2024ApJS..272....2L,2024A&A...687A.177X,2024arXiv240407316L,2024MNRAS.527.1494L,2024MNRAS.52710937Y,2024MNRAS.527.7382A,2024MNRAS.531.2126F}.

\cite{2022ApJ...941...45R} derived the stellar metallicity for a sample of $2$ million giant stars within $30^{\circ}$ of the Galactic center using the \texttt{XGBoost} algorithm. They achieved a remarkably median precision of $\delta\text{[M/H]}<0.1$ by adopting the SDSS DR17 \citep[the Apache Point Observatory Galactic Evolution Experiment, shortened as APOGEE;][]{2022ApJS..259...35A} as the training sample. \cite{2023ApJS..267....8A} applied the \texttt{XGBoost} algorithm to the whole catalog and estimated the stellar atmospheric parameters ($T_\mathrm{eff}, \log g, \mathrm{[M/H]}$) for over 175 million stars. They have a mean stellar parameter precision of 0.1 dex in [M/H], 50 K in $T_\mathrm{eff}$, and 0.08 dex in $\log g$. Besides APOGEE DR17, they also included the metal-poor stars from \cite{2022ApJ...931..147L} in the training sample to break the low-metallicity boundary.

\cite{2023MNRAS.524.1855Z} developed an empirical forward model using a neural network (NN) to estimate the stellar atmospheric parameters, distance, and extinction for 220 million stars from XP spectra. Given the stellar parameters, they could provide a prediction of the XP spectra. \cite{2024arXiv240407316L} developed a novel implementation of a variational auto-encoder which achieves competitive XP spectra reconstructions without relying on stellar labels. Their results suggest that meaningful information related to $[\alpha/\mathrm{Fe}]$ is hidden in the XP spectra. Recently, \cite{2024ApJS..272....2L} and \cite{2024arXiv240401269H} successfully extracted the information on $[\alpha/\mathrm{Fe}]$ for partial XP spectra by using machine learning models trained on APOGEE DR17. In addition, \cite{2024MNRAS.527.1494L} built a transformer-based model and trained it in a self-supervised manner on a compiled data set. Differently from previous studies, they could not only derive stellar parameters from XP spectra, but also predict the XP spectra given the stellar labels. In general, data-driven methods are efficient and accurate in analyzing the XP spectra, and have allowed the deviation of several parameters, like $\mathrm{[\alpha/Fe]}$, that are hard to derive using traditional model-driven methods at such low spectral resolution.

Although data-driven methods have the potential to achieve great success in exploring the low-resolution XP spectra, their output is highly constrained by the quality of the training data. Most studies adopted APOGEE DR17 as a training sample, and several selection criteria were applied to ensure the quality of the stellar labels from APOGEE. Due to those cuts, the $T_\mathrm{eff}$ of the training sample is mainly in the range of $3500$ K to $7000$ K. Therefore, their constructed models are not applicable to stars outside this temperature range. To cover more spectral types in the training sample, \cite{2023MNRAS.524.1855Z} combined the standard AFGK catalog from LAMOST DR8 and the hot stars (high $T_\mathrm{eff}$) catalog from the Hot Payne \citep{2022A&A...662A..66X}. However, there are systematic differences between these two catalogs, and their model prediction exhibits a large scatter from the literature for stars of $T_\mathrm{eff}>7500$ K. Besides the limitation in $T_\mathrm{eff}$, both the APOGEE and LAMOST catalogs have a low boundary of metallicity at $\text{[M/H]} = -2.5$, which prevents the application of the method to more metal-poor stars. Despite the inclusion of more metal-poor stars from other surveys can remove the low-metallicity boundary, the very small size of the training sample prevents an appropriate coverage of the parameter space, and the metallicity estimations only show a very small improvement at $\mathrm{[M/H]}<-2.5$. As a comparison, the high-resolution spectroscopic follow-up of metal-poor candidates selected from the catalogs of $\mathrm{[M/H]_{phot}}$ demonstrated the potential of identifying EMP stars through traditional model-driven methods \cite{2024A&A...687A.177X,2024MNRAS.529L..60M}. 

In this study, we combine data-driven and model-driven methods to construct a catalog ($T_\mathrm{eff}, \log g, \mathrm{[M/H]}$) for about $68$ million stars with XP spectra. To reduce the influence of systematic errors (wiggles), we develop a NN model which corrects the XP spectra according to their stellar labels in \S\ref{sec:method}. This correction helps in overcoming the gap between synthetic spectra and observations, and the stellar atmospheric parameters predicted from the corrected spectra are more reliable, as shown in \S\ref{sec:results}. The paper is summarized in \S\ref{sec:summary}.

\section{Data}\label{sec:data}

We determine and study systematic errors in the Gaia XP fluxes using a sample of XP spectra cross-matched with the final release of APOGEE, included in the DR17 of SDSS. We correct the XP spectra using the systematic pattern detected in the APOGEE sample, and check the impact on the inferred stellar atmospheric parameters (comparing to APOGEE) and on the accuracy of the fluxes \citep[comparing to CALSPEC data;][] {2014PASP..126..711B,2019AJ....158..211B,2022stis.rept....7B}. In this section, we describe each reference catalog in detail. 

\subsection{Gaia XP spectra}

Since XP spectra are described by a set of polynomial coefficients, we use the Python library \texttt{GaiaXPy} to transform these coefficients into flux-calibrated spectra and synthetic photometry. \texttt{GaiaXPy} provides an option to represent the spectrum with a smaller set of basis functions, avoiding the noise associated to higher-order terms. In addition to this truncation, we apply a cut in the wavelength range, eliminating regions where the transmission is less than $10\%$. To remove the influence of the dust extinction, we use the \texttt{Python} package \texttt{extinction}\footnote[4]{\href{https://extinction.readthedocs.io/en/latest/index.html}{https://extinction.readthedocs.io/en/latest/index.html}} adopting the \texttt{ccm89} model \citep{1989ApJ...345..245C}. The reddening for each individual spectrum is calculated from the two-dimensional \texttt{SFD} dust map \citep{1998ApJ...500..525S,2011ApJ...737..103S} using the \texttt{Python} package \texttt{dustmap} \citep{2018JOSS....3..695G}.

The data analysis code \texttt{FER\reflectbox{R}E} \citep{2006ApJ...636..804A} \footnote[5]{\href{https://github.com/callendeprieto/ferre}{https://github.com/callendeprieto/ferre}}, which matches numerical models to the observations by minimizing the $\chi^{2}$, is used to search for the optimal stellar atmospheric parameters ($T_\mathrm{eff}$, [M/H], and $\log g$). Several algorithms are available in \texttt{FER\reflectbox{R}E}, and the Powell’s UOBYQA (unconstrained optimization by quadratic approximation) algorithm \citep{Powell2002UOBYQAUO} is adopted in our work, initializing the search at the grid center. This is a local algorithm, which may lead to incorrect solutions in some cases, but it is chosen due to its high speed, critical for very large samples. We tested applying a Markov chain Monte Carlo algorithm (MCMC, \texttt{algor$=5$} in \texttt{FER\reflectbox{R}E}), which performs a global search, to find out that the results are very similar to those obtained with the algorithm adopted in this work. We therefore conclude that Powell’s UOBYQA algorithm are both efficient and accurate in this case. We computed a fresh library of model spectra with constant resolution using \texttt{Synple}\footnote[6]{\href{https://github.com/callendeprieto/synple}{https://github.com/callendeprieto/synple}}, and Kurucz model atmospheres as in the \texttt{nsc} library of \cite{2018A&A...618A..25A}, spanning $-5 \le$ [M/H] $\le +0.5$, $3500$ K $\le T_{\rm eff} \le 8000$ K, and $1 \le \log g \le 5$. This library is publicly released together with this paper. For the model with variable resolution, the upper limit of $T_{\mathrm{eff}}$ is $12,000$ K.

Gaussian convolution is performed to reduce the resolution of the finely sampled spectral synthesis calculations and match it to the XP data. In this study, we adopt two different resolutions for the model spectra: one is a constant resolution of $R \sim 104$, and the other is a variable resolution that changes with the wavelength as the XP spectra (refer to Table 1 in \citealt{2023A&A...674A...3M}). For the constant resolution model, the corresponding sampled XP spectrum ($S_{\mathrm{const}}$) ranges from $360$ to $990$ nm, and has $330$ points evenly spaced on a logarithmic scale. For the variable resolution, the corresponding XP spectrum ($S_{\mathrm{var}}$) ranges from $360$ to $992.1$ nm, and has $270$ points in increments of the half of Full width at half maximum (FWHM=$\frac{\lambda}{R}$, where $\lambda$ is the wavelength), based on information from \citet[][see also \citealt{2021A&A...652A..86C}]{2023A&A...674A...3M}.

\begin{figure*}
    \centering
    \includegraphics[width = 1.00\textwidth]{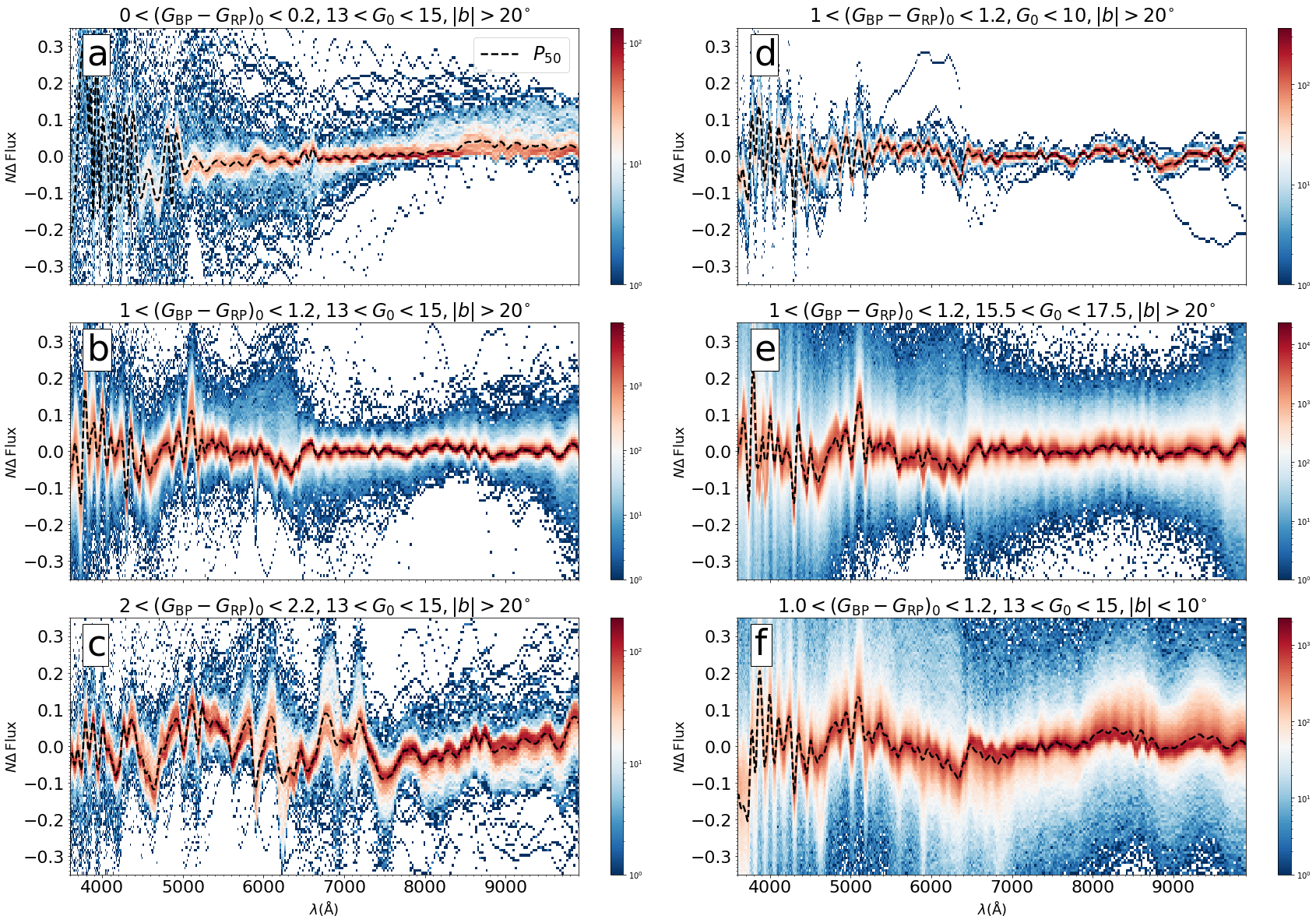}
    \caption{Density distribution of residuals as a function of wavelength for stars with different parameters. Subfigures (a), (b), and (c) are results of stars in bins of the same magnitude range $13<G_{0}<15$ but different stellar colors. Subfigures (b), (d), and (e) show the density distribution of residuals for stars within the same range of color $1.0 < \left( G_{\mathrm{BP}}-G_{\mathrm{RP}} \right)_{0} < 1.2$ but different $G_{0}$. Subfigure (f) shows the residuals of stars of the same color-magnitude as subfigures (b) but in different Galactic latitude ($|b|<10^{\circ}$). The black dash lines represent the $P_{50}$ percentiles distributions, which can be viewed as a robust representation of the patterns of wiggles. From the $P_{50}$ lines we can see that the wiggles change with $\left( G_{\mathrm{BP}}-G_{\mathrm{RP}} \right)_{0}$ and $G_{0}$, but the stellar color has a larger impact than the magnitude. Stars of low latitude have a much more diffuse distribution of $\Delta \mathrm{Flux}$. The large uncertainties of the extinction map towards the Galactic disc may cause a bad extinction correction of $\mathrm{Flux_{XP}}$ for some stars. Therefore, $\mathrm{Flux_{XP}}$ is more likely to deviate from the fitting result $\mathrm{Flux_{fitting}}$ for stars at lower latitude.}
    \label{fig:bin}
\end{figure*}

\texttt{FER\reflectbox{R}E} returns the best fitting spectra that most closely resemble the XP observations. The spectra are normalized by their mean values as in Equation \ref{eq:normalized}, and we define the residual $N\Delta\,\mathrm{Flux}$ as the difference between the normalized fitting ($N\mathrm{{Flux}_{fitting}}$) and XP ($N\mathrm{{Flux}_{XP}}$) spectra.
\begin{equation}
    \begin{aligned}
        &N\mathrm{Flux} = \mathrm{Flux}/\mathrm{Mean(Flux)}\\
            &N\Delta\,\mathrm{Flux} = N\mathrm{Flux_{fitting}}-N\mathrm{Flux_{XP}}
    \end{aligned}
\label{eq:normalized}
\end{equation}
In previous studies of \cite{2023A&A...674A...3M} and \cite{2024ApJS..271...13H}, they also gave a similar definition of the systematic error as $(\mathrm{Flux_{XP}-Flux_{ref}})/\mathrm{Flux_{ref}}$ or $\mathrm{Flux_{ref}}/\mathrm{Flux_{XP}}$, where $\mathrm{{Flux}_{ref}}$ is the corresponding reference spectra from external data. All the fluxes we discuss hereafter are normalized, and the symbol $N$ will be dropped for clarity. The general shapes of the XP spectra are consistent with the best fitting models in most cases. The residuals are presented in a complex function that oscillates with the wavelength as shown in Figure \ref{fig:bin}, and its amplitude is usually larger in the blue band than the red band. In this study, we refer to these residuals as wiggles.

Visual inspection indicates that stars with similar parameters exhibit similar wiggles. To fully understand what are the main parameters that control the pattern, we randomly select $800,000$ stars at different Galactic latitude: $200,000$ at low-latitude ($|b|<10^{\circ}$), $300,000$ at medium-latitude ($10^{\circ} \le |b| < 30^{\circ}$), and $300,000$ at high-latitude ($30^{\circ} \le |b| \le 90^{\circ}$). The corresponding $\mathrm{Flux_{fitting}}$ and $\Delta \mathrm{Flux}$ of these stars are obtained using \texttt{FER\reflectbox{R}E}. The randomly selected XP sources have an extended distribution in the color-magnitude diagram (CMD), with $-0.5 < G_{\mathrm{BP}}-G_{\mathrm{RP}} < 3.5$ and $4<G<17.5$. In \S\ref{sec:Patterns of wiggles} we explore the impact of stellar colors, apparent magnitude, latitude, and metallicity on the wiggles using these randomly selected XP sources.

\subsection{Metal-poor stars sample}\label{sec:metalpoor}

The search for metal-poor stars is one of the most interesting applications of XP data, and therefore we carry out specific evaluations for that type of stars. We select metal-poor stars from three libraries: JINAbase \citep{2018ApJS..238...36A}, the Pristine survey \citep{2019MNRAS.490.2241A}, and the LAMOST-Subaru metal-poor survey \citep{2022ApJ...931..147L,2022ApJ...931..146A}. JINAbase is a collection of chemical abundances and stellar parameters for $1,659$ metal-poor stars ($60\%$ of which have $\mathrm{[M/H]}<-2.5$) from the literature, published in the period between 1991 and 2016. The Pristine survey includes a medium-resolution spectroscopic follow-up of $1,007$ metal-poor candidates identified from the narrow-band photometry, and more than $900$ stars which have been confirmed to have $\mathrm{[M/H]}<-2$. The LAMOST-Subaru survey presents measurements for over $20$ elements in $385$ stars covering a wide metallicity range from $-1.7$ to $-4.3$. After cross-matching with the XP spectra and applying a cut of [M/H] $< -2$, our final metal-poor sample consists of $1,813$ stars. Most of them are at high-latitude, with $E(B-V)<0.1$. The colors and magnitudes of these stars are roughly in the range of $0.5 < G_{\mathrm{BP}}-G_{\mathrm{RP}} < 1.5$ and $6<G<17.5$.

\subsection{APOGEE DR17}

\begin{figure}
    \centering
    \includegraphics[width = 8.8cm]{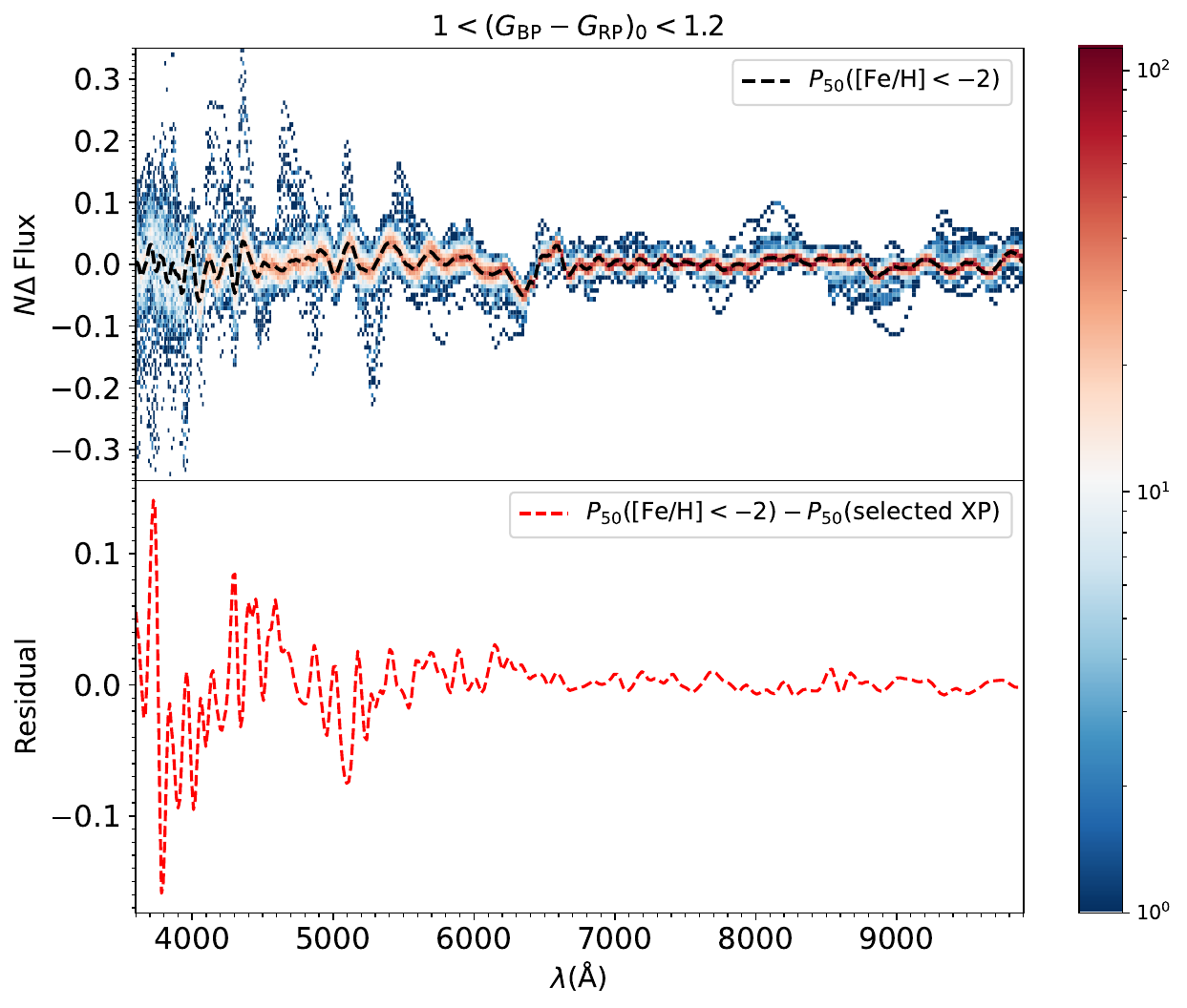}
    \caption{Density distribution of residuals as a function of wavelength for VMP stars. Top panel shows the density distribution of the residuals of VMP stars with $1.0 < G_{\mathrm{BP}}-G_{\mathrm{RP}} < 1.2$ and $13<G<15$. Compared to the relatively more metal-rich sample in Figure \ref{fig:bin}, their wiggles have a smaller amplitude with $\mathrm{Max} \left( |P_{50}| \right) < 0.08$. The bottom panel shows the difference of $\mathrm{P}_{50}$ between these two samples. The main difference is in the blue band of $\lambda<6500$ $\AA$, while the wiggles in the red band is little affected by the change of metallicity.}
    \label{fig:metalpoor}
\end{figure}

APOGEE \citep{2017AJ....154...94M,2019PASP..131e5001W} is a large-scale spectroscopic survey of stars in the Milky Way. APOGEE provides high-resolution ($R=\lambda/\Delta\lambda \sim 22,500$) and high signal-to-noise ratio ($\mathrm{S/N}>70$ typically) spectra throughout the near infrared wavelength range of $1.51-1.70$ $\upmu\mathrm{m}$. 

The stellar atmospheric parameters and elemental abundances are determined by the APOGEE Stellar Parameters and Chemical Abundances Pipeline \citep[ASPCAP;][]{2016AJ....151..144G} based on \texttt{FER\reflectbox{R}E}. APOGEE provides high-precision measurements of the kinematics and chemistry of the Milky Way structures (bulge, disk, and halo). The Seventeenth Data Release of the APOGEE survey contains spectra and abundances for $733,901$ stars \citep{2022ApJS..259...35A}.

\subsection{CALSPEC libraries}

Considering the wide wavelength coverage of the XP spectra, we choose the CALSPEC library \citep{2014PASP..126..711B,2019AJ....158..211B,2022stis.rept....7B} as the reference spectra. CALSPEC is a library of flux standards on the Hubble Space Telescope (HST) system. Most of them have a complete HST's Space Telescope Imaging Spectrograph (STIS) coverage of wavelength from the ultraviolet to near-infrared band with a resolving power $R \approx 560-700$. We find $109$ CALSPEC stars with XP spectra, and their colors and magnitudes are mainly in the range of $-1.0<G_{\mathrm{BP}}-G_{\mathrm{RP}}<1.5$ and $4<G<18$. The CALSPEC collection of spectra are likely among the most accurate flux calibrated spectra available, with a $2-3\%$ accuracy in absolute flux. The uncertainty of the monochromatic flux at $555.75$ nm ($555.6$ nm in air) is $0.5\%$ or $0.005$ mag (see \citealt{2014PASP..126..711B}, and the CALSPEC Calibration Database).

\begin{figure*}[htbp!]
    \centering
    \includegraphics[width = 1.0\textwidth]{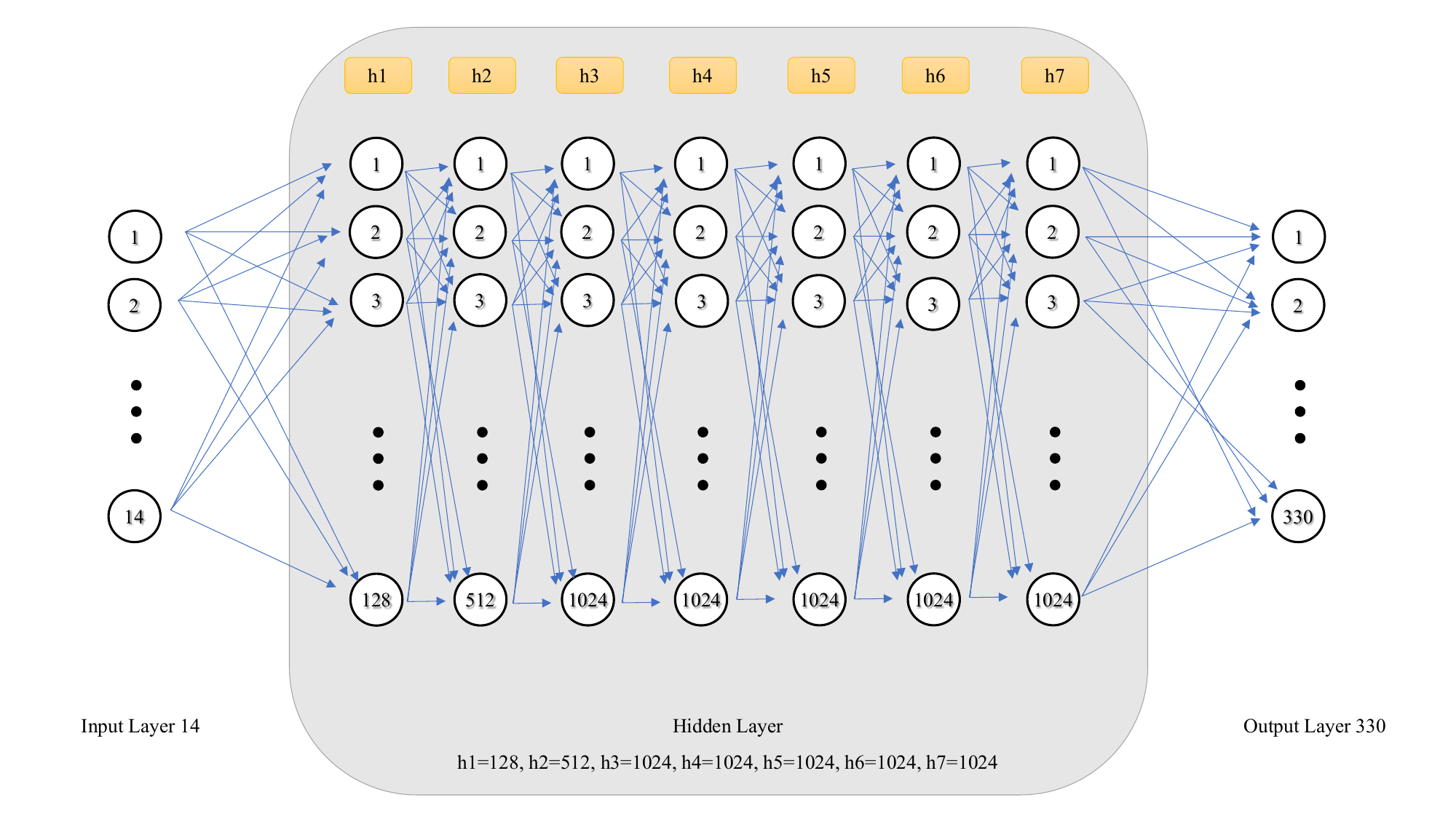}
    \caption{Diagram of our NN model based on \texttt{Pytorch}.}
    \label{fig:NNmodel}
\end{figure*}

\subsection{LAMOST}

LAMOST is a ground-based innovative telescope designed with a large aperture and a wide field-of-view \citep{2006ChJAA...6..265Z,2012RAA....12..723Z,2012RAA....12.1197C,2012RAA....12.1243L}. Beginning with a pilot survey in 2011, LAMOST has observed over ten million spectra in the Northern sky with a limiting magnitude of $r=17.8$ \citep{2022Innov...300224Y}. 

LAMOST Low-Resolution Spectroscopic (LRS) survey provides low-resolution ($R \sim 1,800$) spectra with a wavelength coverage of $0.37-0.90$ $\upmu\mathrm{m}$. In this study, we use the LRS Stellar Parameter Catalog of A, F, G and K stars from the LAMOST DR11, which contains the atmospheric parameters of $7,774,147$ stars with a typical error of $\sim 43$ K for $T_\mathrm{eff}$, $0.06$ dex for $\log g$, and $0.04$ dex for [M/H].

\section{Pattern of wiggles}\label{sec:Patterns of wiggles}

Comparison between XP spectra and model predicted flux, \cite{2023A&A...674A...3M} shows that the residuals are correlated with $G_{\mathrm{BP}}- G_{\mathrm{RP}}$ color and $G$ magnitude. The stellar color is closely associated with surface temperature, which plays a critical role in shaping the spectra energy distribution (SED) through photoionization \citep{2023Atoms..11...61A}, mainly through hydrogen atoms and the H$^-$ ion. The dependence of the wiggles on magnitude may likely be caused by the complex internal calibration of XP spectra. As illustrated by \cite{2023A&A...674A..27A}, the spectra of stars brighter than $G$ = 11.5 are recorded in 2D windows, which have different read-out configuration (gates) depending on magnitude, while fainter stars are mainly observed as 1D windows and under a more uniform readout strategy. \cite{2023A&A...674A...3M} noted that a built-in assumption in the external calibration (which involves calibrating the internally calibrated, continuously represented mean spectra to an absolute system) is that the instrument model is independent of brightness (or magnitude). Therefore, it cannot easily account for inconsistencies caused by the different observing modes, which may finally result in a correlation between residuals and $G$ magnitude.

Inspired by previous studies, we randomly distribute our selected XP sources, limited to $E(B-V)<2$, into different bins in de-reddened  $\left( G_{\mathrm{BP}}-G_{\mathrm{RP}} \right)_{0}$ color and  $G_{0}$ magnitude. Figure \ref{fig:bin} shows the density distribution of residuals as a function of wavelength for stars with different values of $\left( G_{\mathrm{BP}}-G_{\mathrm{RP}} \right)_{0}$, $G_{0}$, and Galactic latitude $b$. We find that the wiggles change dramatically with color. For stars in the same color bins, we can see some scatter, but most of them have a very similar pattern, especially for stars in the range $1.0 < \left( G_{\mathrm{BP}}-G_{\mathrm{RP}} \right)_{0} < 1.2$. 

Panels (b), (d), and (e) of Figure \ref{fig:bin} show the distributions of residuals for stars with the same color but different magnitude $G_{0}$. The residuals' pattern clearly changes with $G_{0}$ as well, although to a lesser extent than with $(G_{\mathrm{BP}}-G_{\mathrm{RP}})_{0}$. The stellar color is a major determining factor of the SED, while the apparent magnitude $G_{0}$ is connected to the SED through the instrument model. Since the wiggles in the residuals largely depend on the SED, it is only natural that the stellar color plays a much more important role. For stars in the same color-magnitude bin but located at lower Galactic latitudes, the general pattern is similar, but the residuals have a much more diffuse distribution. We conclude that the stellar color $G_{\mathrm{BP}}-G_{\mathrm{RP}}$, apparent magnitude $G$, and dust extinction $E(B-V)$ should be considered when modeling the observed systematic patterns.

\cite{2024ApJS..271...13H} also considered the intrinsic color and extinction in their proposed corrections, but disregarded the weaker effects of [M/H] and $\log g$. This choice simplifies the correcting process, focusing on the accuracy of the correction for most stars. However, the libraries they used mainly consist of stars with $\mathrm{[M/H]}>-1$, and the SEDs of very metal-poor (VMP) stars are quite different from those of metal-rich ones. To fully understand the influence of [M/H], we adopt the sample from \S\ref{sec:metalpoor}, and obtain its density distribution of residuals in Figure \ref{fig:metalpoor}. We can see that the largest differences are concentrated on the blue band, where the residuals of VMP stars have a smaller amplitude, with the maximum absolute value $\mathrm{Max} \left( |P_{50}| \right) < 0.08$. In general, the pattern of fitting residuals for VMP stars is far more featureless than for metal-rich stars. Since one of our aims is to make an accurate estimation of metallicity from the wiggle-corrected XP spectra, we should take extra care with the influence of [M/H].

\section{Method}\label{sec:method}

Our aim is to correct the wiggles observed in the residuals for a given spectrum and calculate atmospheric parameters by fitting the observations with model spectra using \texttt{FER\reflectbox{R}E}. Therefore, we first need to characterize the systematic patterns as a function of color, magnitude, reddening and metallicity. This is what machine learning is good at. Therefore, we use the main relevant parameters as input into a neural-network model with multiple hidden layers, where the residuals from the fits are the output. We may not have the information of metallicity for a given star. Therefore, we strongly prefer to avoid using metallicity as input, and instead adopt various metallicity-sensitive photometric indices.

It is important to stress that there is a limited number of spectra with metallicity below $-2.5$ to train the NN model, so we do not trust and do not apply the derived correction to stars with metallicity under $-2.5$. In this section, we first describe how we predict the patterns via the NN, including the training dataset and the architecture of our NN model. Then we provide additional details on how we correct and fit all XP spectra in Gaia DR3.

\subsection{Training database}

To ensure the patterns in the residuals we characterize really reflect the differences between the XP and the theoretical spectra, we only use spectra with reliable estimations of atmospheric parameters, judging from the differences with the APOGEE parameters. We refer to the resulting sample as the training APOGEE sample (TAS) hereafter. The APOGEE sample is firstly cleaned by applying the cuts below, where the symbol A indicates APOGEE parameters:
\begin{itemize}
    \setlength{\itemsep}{7pt}
    \item S/N $>$ 70;
    \item $\sigma_{T_{\mathrm{eff}}^{\mathrm{A}}} \le 250$, $\sigma_{\log g^{\mathrm{A}}} \le 0.5$;
    \item $3500 \le T_{\mathrm{eff}}^{\mathrm{A}} \le 8000$ K or $\le 12000$ K, $1 \le \log g^{\mathrm{A}} \le 5$.
\end{itemize}

We adopted two different values as the upper limits of $T_{\mathrm{eff}}^{\mathrm{A}}$ for the two models with different resolutions. The upper limit is $8000$ K for $S_{\mathrm{const}}$ and $12000$ K for $S_{\mathrm{var}}$. In the APOGEE DR17 catalog, there are some sources with Gaia $\texttt{source\underline{ }id}=0$, which are dropped in our APOGEE sample. In addition, for sources with the same Gaia \texttt{source\underline{ }id}, we only keep the first one ($\sim 10\%$ sources are duplicated). Then we obtain the initial parameters for the APOGEE sample (IAS). Next, we calculate the atmospheric parameters for XP spectra using \texttt{FER\reflectbox{R}E}. After deriving $T_{\mathrm{eff}}$, $\log g$, and [M/H], we trim down the IAS applying the following constraints:
\begin{itemize}
    \setlength{\itemsep}{7pt}
    \item $|T_{\mathrm{eff}}^{\mathrm{XP}}-T_{\mathrm{eff}}^{\mathrm{A}}| < 200$
    \item $|\log g^{\mathrm{XP}} - \log g^{\mathrm{A}}| < 0.5$
    \item $|\mathrm{[M/H]^{\mathrm{XP}}} - \mathrm{[M/H]}^{\mathrm{A}}| < 0.5$,
\end{itemize}
where $T_{\mathrm{eff}}^{\mathrm{XP}}, \log g^{\mathrm{XP}}, \mathrm{[M/H]^{\mathrm{XP}}}$ are parameters estimated from the Gaia XP spectra by \texttt{FER\reflectbox{R}E}. Finally, our TAS includes $157,478$ stars for $S_{\mathrm{const}}$, and $131,173$ stars for $S_{\mathrm{var}}$.

\subsection{Input and output}

As mentioned before, the shape of the residuals is highly related to the color, magnitude, and reddening of a star, and to a lesser extent, to its metallicity. Therefore, we use various of colors and magnitudes, including metallicity-sensitive colors, generated from XP spectra, as input for our NN model. We employ \texttt{GaiaXPy} and the built-in photometric system \citep{2023A&A...674A..33G} to obtain the SkyMapper magnitudes $u$, $v$, $g$, $i$, and the metallicity-sensitive colors \citep{2021ApJS..254...31C} $g-i$, $v-g-0.9\times(g-i)$, and $u-v-0.9\times(g-i)$. In addition, the reddening $E\left(B-V\right)$ calculated from \texttt{dustmaps} is included as one of the parameters.

To summarize, the $14$ parameters involved are: 
\begin{itemize}
    \item metallicity-sensitive colors: $g-i$, $v-g-0.9\times(g-i)$, $u-v-0.9\times(g-i)$
    \item SkyMapper photometric passbands: $u$, $v$, $g$, $i$
    \item Gaia photometric passbands and colors: 
    \texttt{phot\underline{ }g\underline{ }mean\underline{ }mag}, 
    \texttt{phot\underline{ }bp\underline{ }mean\underline{ }mag}, 
    \texttt{phot\underline{ }rp\underline{ }mean\underline{ }mag}, 
    \texttt{bp\underline{ }rp}, \texttt{bp\underline{ }g}, 
    \texttt{rp\underline{ }g}
    \item reddening: $E\left(B-V\right)$.
\end{itemize}
The output is the predicted flux corrections as a function of wavelength.

\subsection{Neural network architecture}

A simple NN model based on \texttt{Pytorch} \citep{pytorch,NEURIPS2019_bdbca288}\footnote[7]{\href{https://github.com/pytorch/pytorch/tree/main}{https://github.com/pytorch/pytorch/tree/main}} is built for training with the TAS. The basic diagram illustrating our NN model is summarized in Figure \ref{fig:NNmodel}. Briefly, we have $7$ hidden layers in our model, and the number of neurons per hidden layer is shown in the figure. We have $14$ input elements, as described above. The length of output for each target is $330$ (for $S_{\mathrm{const}}$, and $270$ for $S_{\mathrm{var}}$). The loss function and optimizer adopted in our model are \texttt{MSELoss} and \texttt{Adam}. The initial learning rate for training is set to $0.001$, and it is reduced to half that value after $15$ epochs, with the lower limit set to $10^{-6}$. In order to prevent overfitting, we use \texttt{EarlyStopping} in our model.

The TAS data set is divided into a training sample ($60\%$), a validation sample ($20\%$), and a testing sample ($20\%$). When the loss curve for the validation sample is not declining after $25$ epochs, the train process is stopped. Our finally adopted model is trained on an Nvidia RTX 4090 in about 5 minutes for $234$ epochs.

\begin{figure}[htbp!]
    \centering
    \includegraphics[width = 8.8cm]{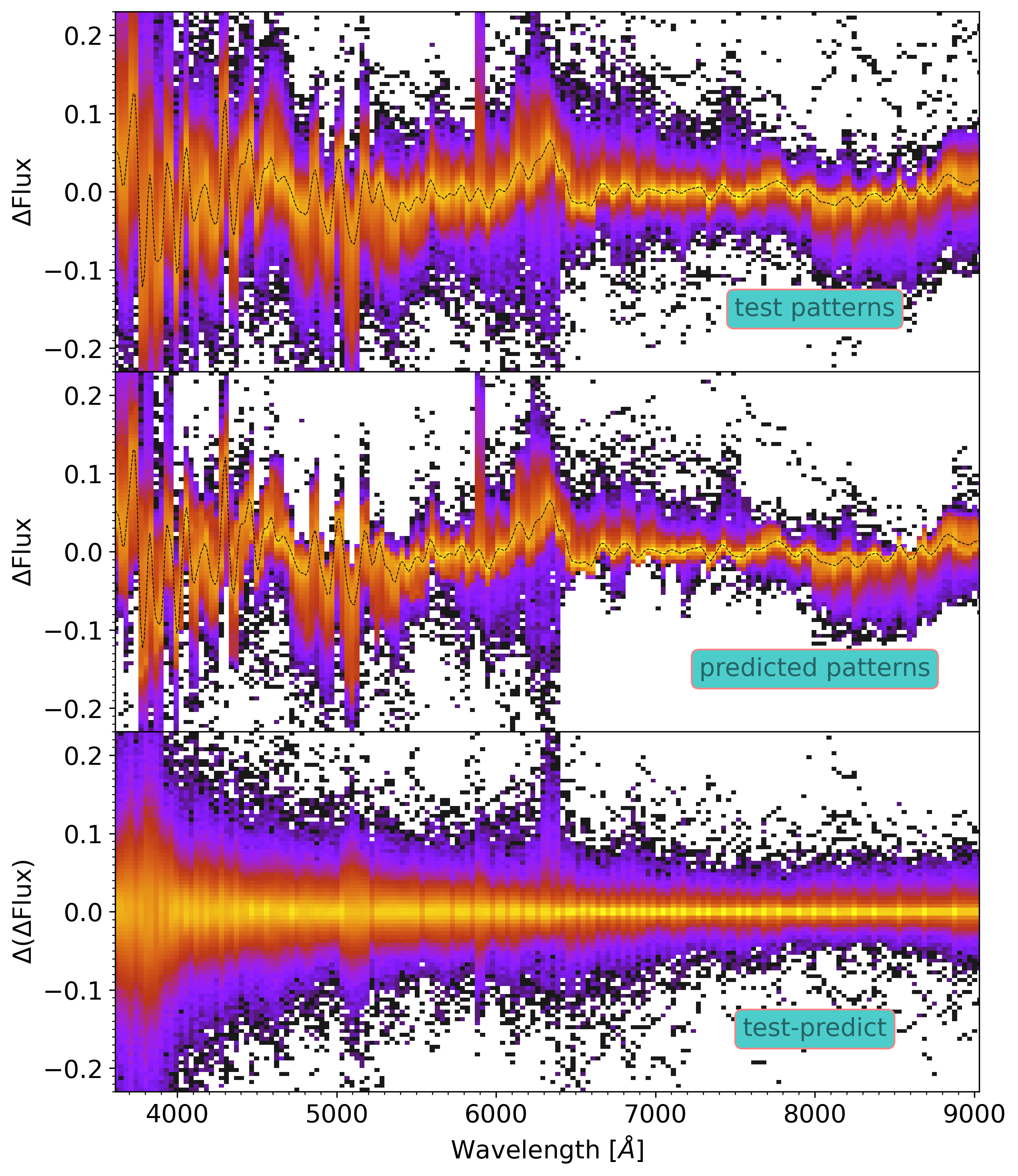}
    \caption{Comparison between the real patterns from TAS ($20\%$ of the whole sample, top panel) and the NN model predicted patterns (middle panel). The bottom panel shows the difference between the real and predicted patterns, from which we can see that the wiggles disappear.}
    \label{fig:NNoutput}
\end{figure}

\begin{figure*}[t!]
    \centering
    \includegraphics[width = 1.00\textwidth]{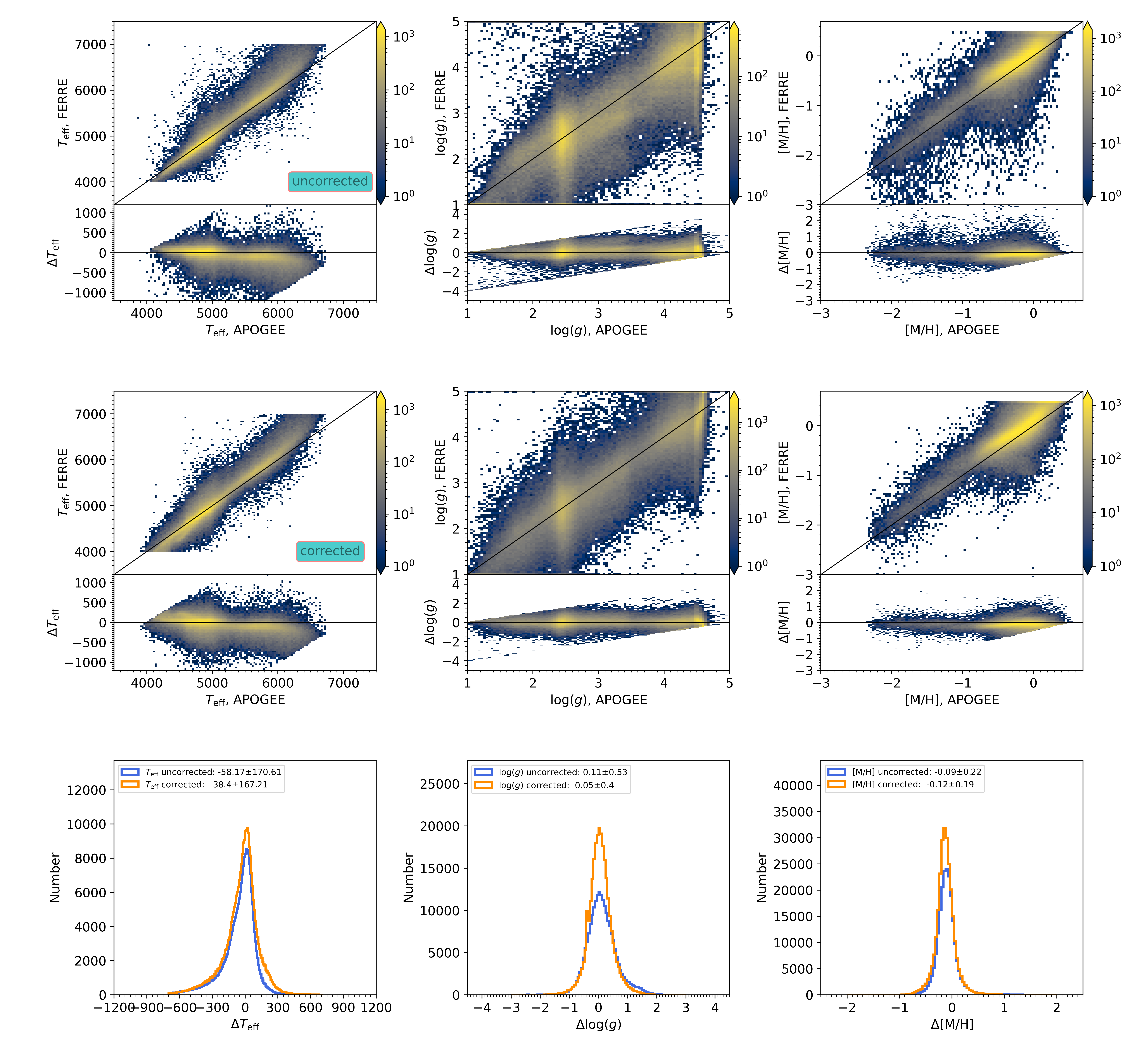}
    \caption{Testing on atmospheric parameters using sampling $S_{\mathrm{const}}$. Top panels: Comparison of $T_{\mathrm{eff}}$, $\log g$ and [M/H] estimated from original XP spectra (Y-axis) and APOGEE survey (X-axis). Middle panels: Similar to the top panels, but with Y-axis replaced by the results from corrected XP spectra. The color bar in each subfigure displays the number density. Bottom panels: Histograms present the differences of $T_{\mathrm{eff}}$, $\log g$ and [M/H] between XP and APOGEE before and after correcting the systematic patterns. The mean values and standard deviations are shown in the labels.}
    \label{fig:apogeecom1}
\end{figure*}

Our NN model performs very well in predicting the pattern in the residual for a given star. In Figure \ref{fig:NNoutput}, we demonstrate the predicted wiggles for $20\%$ sources in the TAS. As we can see from the figure, the predicted wiggles show significantly less scatter than the actual wiggles, but the mean pattern in the actual wiggles is clearly shown in the predicted ones. The bottom panel of this figure also indicates the patterns are largely removed after applying the correction.

\subsection{Fit all XP spectra} \label{sec:fit_allspectra}

To determine atmospheric parameters for all the stars with Gaia XP spectra, we use \texttt{FER\reflectbox{R}E}. This process is performed in three steps:
\begin{itemize}
    \item[(1)] We fit all de-reddened XP spectra with \texttt{FER\reflectbox{R}E} and arrive at an approximate estimation of $T_{\mathrm{eff}}$ and [M/H]:
    \begin{itemize}
        \item Stars with extinction $A_{\mathrm{V}} > 15$ are removed; (This value is set quite high, as we do not intend to exclude any spectrum for its high reddening. The limit is primarily in place to prevent calculation errors. If we lower this value to $1.0$ or $2.0$, it will not exclude many more stars, as more than $99.2\%$ of the stars in our final catalog have $A_{\mathrm{V}}<2.0$.)
    \end{itemize}
    \item[(2)] We use the NN model to correct the systematics in the spectral energy distribution (wiggles):
    \begin{itemize}
        \item Spectra for which we are not able to generate the full set of synthetic photometry are not corrected (The reason is explained later);
        \item Spectra with $T_{\mathrm{eff}}$ in the first estimation larger than $8000$ K will not be corrected (only activated during the process of fitting spectra with variable resolution models);
        \item Spectra with $\mathrm{[M/H]} \le -2.5$ in the first estimation will not be corrected.
    \end{itemize}
    \item[(3)] We fit the XP spectra with \texttt{FER\reflectbox{R}E} again.
\end{itemize}
Some spectra can not produce all the passbands we require, typically the $u$ and $v$ bands of SkyMapper in our test, with the $u$ band being the most affected. According to Table 1 in \cite{2024PASA...41...61O}, the central wavelength and filter FWHM for the $u$ band are $350$ nm and $43$ nm, respectively, while for the $v$ band, these values are $384$ nm and $31$ nm. The starting wavelength of XP spectra is $330$ nm, covering most wavelength range of the $u$ band and the full wavelength range of the $v$ band. However, \texttt{GaiaXPy} fails to generate these two bands for some spectra, primarily for faint stars, though not all faint stars are affected by this issue.

\section{Results}\label{sec:results}

In this section, we discuss the estimation of parameters for the IAS after correcting the systematic pattern predicted by our NN model trained on the TAS data set. As described below, this correction makes the XP spectra fit the model spectra better, and in particular, we find a significantly better agreement with the HST CALSPEC observations.

\subsection{Stellar atmospheric parameters}\label{sec:parameters}

Besides determining the absolute flux, we find that the correction of the systematic pattern can improve the estimation of stellar atmospheric parameters from XP spectra. We first calculate parameters from XP spectra with sampling $S_{\mathrm{const}}$. In Figure \ref{fig:apogeecom1} we show the distributions of the stellar atmospheric parameters ($T_\mathrm{eff}$, $\log g$, [M/H]) inferred from the original and wiggle-corrected spectra, comparing with parameters from the APOGEE survey. The top and middle panels compare parameters from XP and APOGEE before and after the correction. We keep stars with $4000 \le T_{\mathrm{eff}} \le 7000$ K to compare with APOGEE, because our analysis for stars with lower and higher temperatures performs worse. In addition, the stars in this figure have passed the selection based on the quality flag \texttt{dflux\underline{ }per}, which describes the percentage of data points from $\Delta\mathrm{Flux} \equiv \mathrm{Flux_{XP}-Flux_{model}}$ exceeding $\pm0.05$. We retain stars with \texttt{dflux\underline{ }per} below $20\%$ before the correction and $8\%$ after the correction. It is worth noting that more stars remain after correction, as the correction improves the fit between the XP spectra and the models. As a result, even with more stringent selection criteria, we still obtain more stars than before the correction.

Figure \ref{fig:apogeecom1} shows that our estimations are highly consistent with APOGEE's parameters both before and after corrections within the range $4000 \le T_{\mathrm{eff}} \le 7000$ K. But comparing the middle panels with the top panels, one can easily find that estimations for $T_\mathrm{eff}$, $\log g$ and [M/H] become somewhat better after correction, especially when examining the differences in $\log g$ and [M/H]. Correcting the systematic pattern not only reduces the biases in $T_{\mathrm{eff}}$ and $\log g$, but it also makes our results more precise, as suggested from the smaller dispersion relative to APOGEE results.

We provide below a brief summary of the changes in the mean value and standard deviation of $\Delta T_{\mathrm{eff}}$, $\Delta \log g$, and $\Delta \mathrm{[M/H]}$ for the nominal $S_{\mathrm{const}}$ analysis:
\begin{itemize}
    \item $\Delta T_{\mathrm{eff}}$: from $-58.17 \pm 170.61$ to $-38.40 \pm 167.21$;
    \item $\Delta \log g$: from $0.11 \pm 0.53$ to $0.05 \pm 0.40$;
    \item $\Delta \mathrm{[M/H]}$: from $-0.09 \pm 0.22$ to $-0.12 \pm 0.19$.
\end{itemize}
The most significant improvement is in the estimation of $\log g$, where the dispersion $\sigma_{\Delta \log g}$ is reduced from $0.53$ to $0.40$.

We also calculate parameters for $S_{\mathrm{var}}$ for the same sample of stars (IAS) with the same selection rules. We find that $\Delta T_{\mathrm{eff}}$, $\Delta \log g$, and $\Delta \mathrm{[M/H]}$ are very similar with previous results using sampling $S_{\mathrm{const}}$, summarized as follows.
\begin{itemize}
    \item $\Delta T_{\mathrm{eff}}$: from $-62.11 \pm 177.21$ to $-50.28 \pm 163.46$;
    \item $\Delta \log g$: from $-0.09 \pm 0.73$ to $-0.05 \pm 0.47$;
    \item $\Delta \mathrm{[M/H]}$: from $-0.13 \pm 0.23$ to $-0.14 \pm 0.19$.
\end{itemize}
Despite the overall results are quite similar for the two different samplings, those for $S_{\mathrm{const}}$ before correcting patterns are slightly better. After correcting the patterns, we find that $\Delta T_{\mathrm{eff}}$, $\Delta \log g$, and $\Delta \mathrm{[M/H]}$ from the two samplings are similar. However, the constant sampling still seems to provide better results, especially for $T_{\mathrm{eff}}$ and $\log g$. Therefore, we adopt constant sampling in the following sections and our series of papers.

\begin{figure}[t!]
    \centering
    \includegraphics[width = 8.8cm]{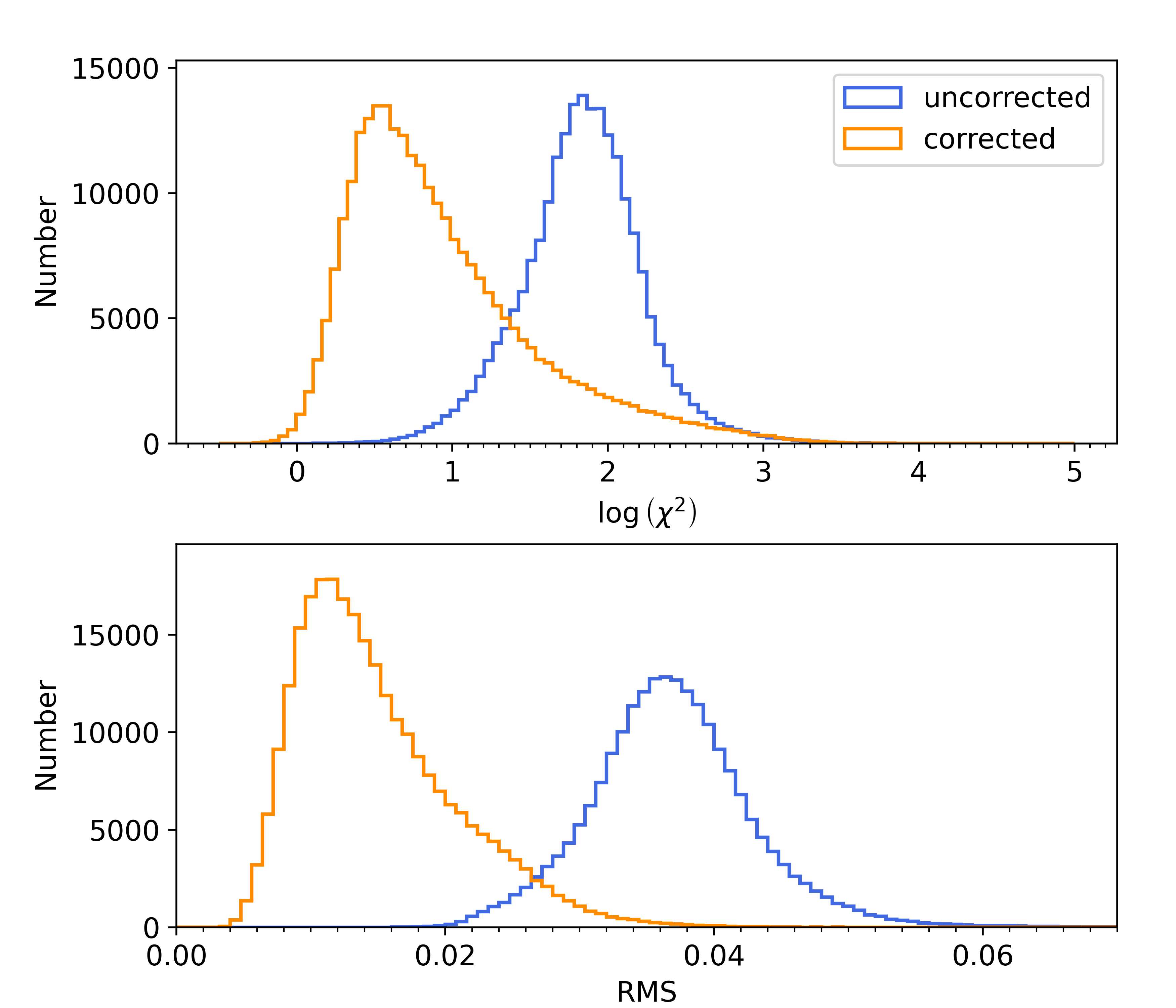}
    \caption{Histograms present $\log_{10} \left( \chi^{2} \right)$ and RMS between the XP spectra and model spectra for the IAS data set before and after correcting the pattern, indicating by blue and orange, respectively.}
    \label{fig:rms}
\end{figure}

\subsection{Flux corrections}

Correcting the wiggles has a significant impact on the fitting of XP spectra. We examine the $\log_{10} \left( \chi^{2} \right)$ and RMS (root mean square) distributions and both parameters are significantly reduced, as shown in Figure \ref{fig:rms} for the stars in the IAS. The RMS is defined:
\begin{eqnarray}
    \mathrm{RMS} =  \sqrt{\frac{1}{n} \sum_{i} \left( F^{\mathrm{XP}}_{i} - F^{\mathrm{model}}_{i} \right)^{2}}
\end{eqnarray}
where $F^{\mathrm{XP}}, F^{\mathrm{model}}$ are the fluxes of XP spectrum and model spectrum, $i$ indicates the points of sampling, and $n$ is the number of data points. Our correction makes $\log_{10} \left( \chi^{2} \right)$ and RMS becoming way much smaller. As shown in the bottom panel of Figure \ref{fig:rms}, the peak of the RMS histogram decreases from $3.7\%$ to $1.2\%$. In the following we check whether the correction helps in bringing the XP spectra closer to the CALSPEC data, which have a superb flux calibration.

\subsection{The CALSPEC library} 

Given the excellent absolute flux calibration of the CALSPEC spectra, we use this library to perform an independent check of our proposed corrections. Since they have significantly larger resolving power, we apply Gaussian convolution to match their resolution to the XP spectra (and our model spectra). Then, we perform an interpolation of the smoothed spectra to match the sampling of XP data, and correct extinction in the same way we do for the XP spectra. Only spectra of stars estimated to have $T_{\mathrm{eff}}<10^{4}$ K are used in this comparison, based on the parameter \texttt{teff\underline{ }gspphot} in Gaia DR3.

\begin{figure}[htbp!]
    \centering
    \includegraphics[width = 8.8cm]{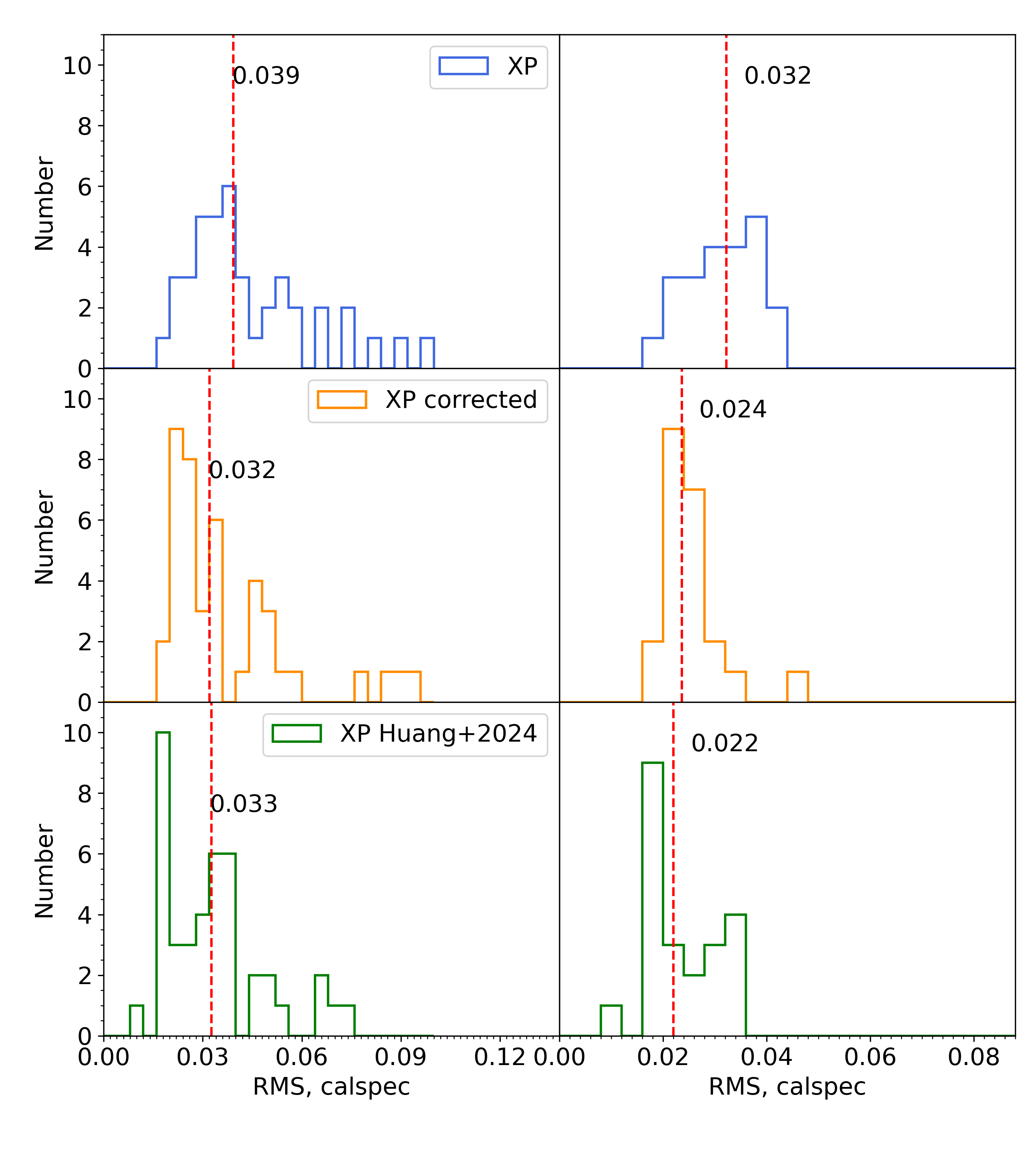}
    \caption{Distribution of RMS between CALSPEC and XP spectra. Left panels: Histograms present RMS between the XP spectra and the spectra from CALSPEC libraries before and after correcting patterns, indicating by blue and orange, respectively. The green histogram shows the corresponding results obtained using the \texttt{Python} package from \cite{2024ApJS..271...13H} to correct the XP spectra. Right panels: Similar to those panels in the left, but with stars that have passed the quality cuts applied in \S\ref{sec:parameters}. The red dashed lines represent the median values, which are also indicated by the number displayed in each panel.}
    \label{fig:calspec_rms}
\end{figure}

\begin{figure}[htbp!]
    \centering
    \includegraphics[width = 8.8cm]{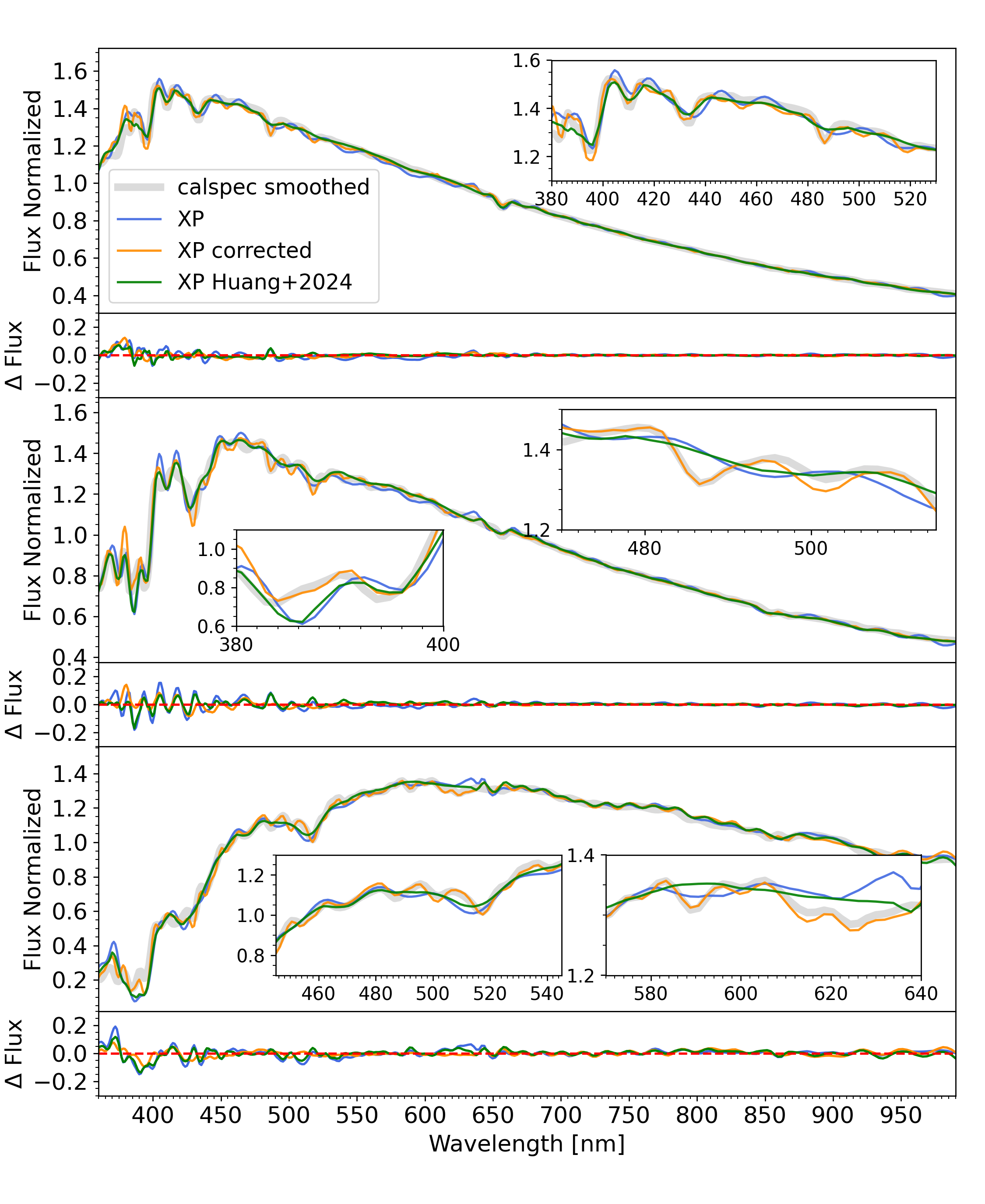}
    \caption{Examples of how pattern correction can help improve the fitting between the XP spectra and the CALSPEC. Three spectra with different temperatures are presented in the sub-panels, from top to bottom: BD+54 1216, HD 115169, and KF06T2 in CALSPEC. In each panel, the smoothed CALSPEC spectrum is depicted as a bold gray line, while the XP spectra, with and without correcting patterns, are shown in blue and orange, respectively. The green lines show the corrected spectra using the package provided by \cite{2024ApJS..271...13H}. To enhance readability, we include a few zoomed-in diagrams in each panel to highlight the improvements on spectrophotometry. The $\Delta\mathrm{Flux}$ between XP and CALSPEC are also presented at the bottom of each panel.}
    \label{fig:examples_calspec}
\end{figure}

\begin{figure*}[htbp!]
    \centering
    \includegraphics[width = 1.00\textwidth]{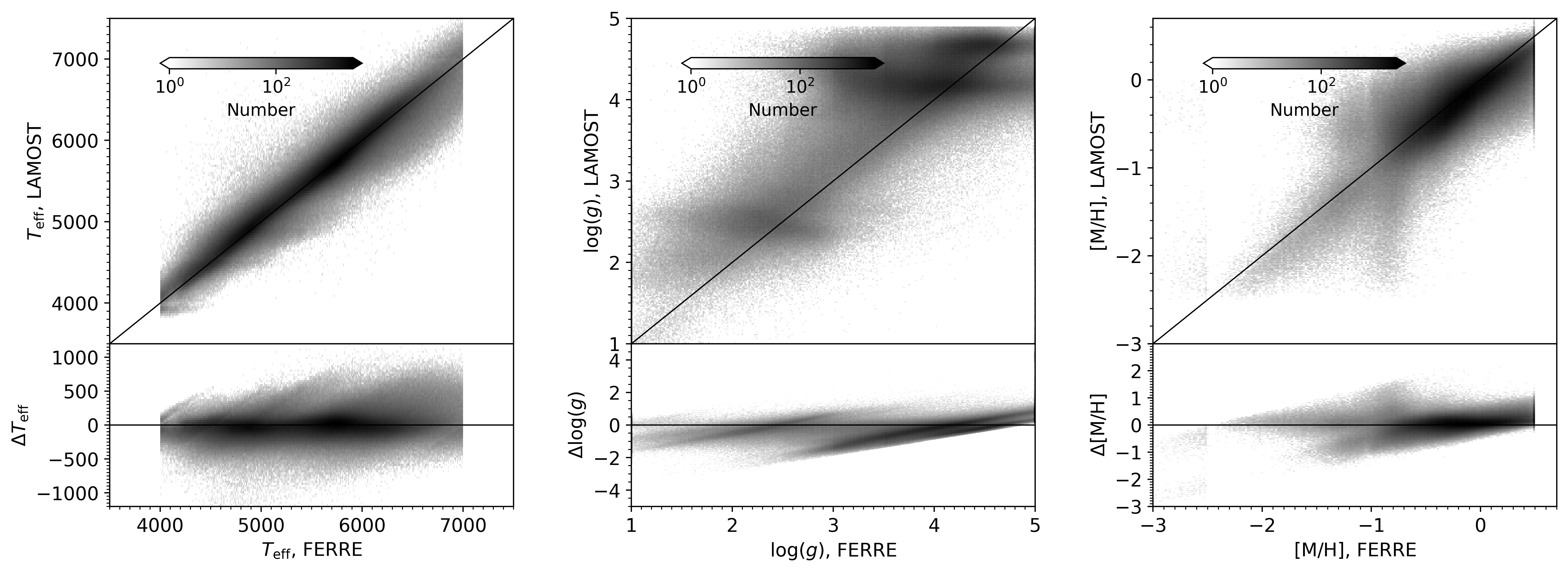}
    \caption{Comparison of atmospheric parameters $T_{\mathrm{eff}}$, $\log g$, and [M/H] between our catalog and LAMOST DR11 low-resolution catalog.}
    \label{fig:compare_lamost}
\end{figure*}

We first check the RMS between the XP spectra and the spectra from CALSPEC, comparing the results before and after the correction. All spectra from XP and CALSPEC are de-reddened with the same method (see in \S\ref{sec:data}). The results for stars without quality cuts are shown in the left-hand panels of Figure \ref{fig:calspec_rms}, and the right panels display the results for stars that have passed the quality cuts. We find that the RMS with the correction of the pattern is overall reduced, with only one exception for stars that have passed the quality cuts. From the median RMS values, indicated by the red dashed lines in the figure, our correction reduces the RMS from $3.9\%$ to $3.2\%$, or from $3.2\%$ to $2.4\%$, depending on whether quality cuts are applied. Overall, the XP spectra are fit better with CALSPEC after correcting the systematic pattern. We also use the package provided by \cite{2024ApJS..271...13H} to correct XP spectra as a comparison, shown in the two subfigures in the bottom panels. We de-redden and interpolate the spectra corrected by the package to match the sampling used in this paper, because the default sampling in our work differs from that of \cite{2024ApJS..271...13H}. From the orange and green histograms, shown in the middle and bottom panels in Figure \ref{fig:calspec_rms}, the overall improvements from our correction and the correction by \cite{2024ApJS..271...13H} are comparable, regardless of whether quality cuts are applied. We should also mention that the resolution of our model spectrum is not exactly the same as that of the XP spectrum, as we use a constant resolution of around $100$. However, the typical resolution of XP spectra is below $\sim 100$. Consequently, the XP spectrum may be over-corrected, with some wiggles potentially arising from the higher resolution of models adopted in this paper. However, we have checked the results using variable resolution and reach the same conclusions.

We also present plots for a few individual XP spectra comparing them to CALSPEC in Figure \ref{fig:examples_calspec}. After correcting the pattern, the XP spectra match better the CALSPEC data smoothed to the resolution of the Gaia spectrophotometry. Nonetheless, there are some wavelengths at which the corrected version gives poorer agreement with CALSPEC. It is worth noting that the resolution adopted in \cite{2024ApJS..271...13H} is lower than that used in this study, which accounts for some of the observed differences between the corrected spectra by \cite{2024ApJS..271...13H} and CALSPEC at the blue end of the spectra. We have also repeated the check with the Next Generation Spectral Library \citep[NGSL;][]{2006hstc.conf..209G,2023ApJS..266...41P}, but the improvement is not so clear, which we associate to the lower quality of the flux calibration of the library compared to the exquisite CALSPEC data.

\subsection{Atmospheric parameters catalog of Gaia XP spectra}

In \S\ref{sec:fit_allspectra} we describe our analysis of the whole sample of Gaia XP spectra. Using the same restrictions in \S\ref{sec:parameters} ($4000 \le T_{\mathrm{eff}} \le 7000$ K, and a maximum of $8\%$ or $20\%$ of the data points with $\Delta\mathrm{Flux}$ larger than $0.05$), we build our catalogs of stellar parameters for stars with Gaia XP spectra. We have a global catalog containing $68,394,431$ stars, and a metal-poor catalog containing $124,188$ stars that have $\mathrm{[M/H]}<-2.5$ and $A_{\mathrm{V}} \le 1.5$.

\subsubsection{Comparison with LAMOST}

The global catalog has also been cross-matched with LAMOST DR11 low-resolution catalog using the Gaia \texttt{source\underline{ }id}, finding $3,006,606$ common sources. We compare the atmospheric parameters $T_{\mathrm{eff}}$, $\log g$, and [M/H] between our results and LAMOST in Figure \ref{fig:compare_lamost}.

\subsubsection{Comparisons with XP catalogs from the literature}

We also compare our results with recent catalogs from literature. The details of the comparisons with \cite{2023ApJS..267....8A} and \cite{2023MNRAS.524.1855Z} are provided in Appendix \ref{appendix_a}. The bottom line is that there is fair agreement between the three catalogs, suggesting that the uncertainties in our effective temperatures and metallicities are similar to those in the other catalogs, typically about $150$ K and $0.2$ dex, respectively, while our gravities are more uncertain than those in the other catalogs, about $0.4$ dex for us but nearly 0.2 dex for the others. The analysis by \cite{2023ApJS..267....8A} and \cite{2023MNRAS.524.1855Z} are different in nature from ours. Not only they employ data-driven methods, while our results are obtained from synthetic spectra based on model atmospheres, but they make use of the trigonometric parallaxes from Gaia, while we do not. We plan to adapt our algorithm to make use of that valuable information, which can directly and dramatically improve the retrieved surface gravities, as well as constraining better the effective temperatures and metallicities. Nonetheless, our present results are independent from models of stellar structure and evolution.

The most used training datasets for machine learning, such as the catalogs from APOGEE and LAMOST, have a fairly restrictive lower boundary in metallicity, which limits their application to detect metal-poor star candidates. Additional comparisons with \cite{2023ApJS..267....8A} and \cite{2023MNRAS.524.1855Z} regarding VMP stars are also presented in Appendix \ref{appendix_a}. In summary, our catalog offers some advantages in the parameter range beyond those covered by previous work, and has more clean and direct connection to physical models of stellar atmospheres.

\subsubsection{Star clusters}

We use open clusters (OCs) and globular clusters (GCs) to validate the metallicities in our catalog. Details of the comparison are outlined in Appendix \ref{appendix_b}. Although there are some outliers in the metallicity distribution, as seen in GC NGC 3201, the overall results are in excellent agreement with the literature.

\subsubsection{The final catalog}

The first column in our global catalog is \texttt{name}, representing the Gaia \texttt{source\underline{ }id} of each source. Following this, we provide the atmospheric parameters $T_{\mathrm{eff}}$, $\log g$, and [M/H] as \texttt{Teff}, \texttt{logg}, \texttt{FeH}. The $\log_{10} \left( \chi^{2} \right)$ value for each spectrum is also included in the column \texttt{log10\underline{ }chi2}. Another parameter reflecting the quality of the atmospheres parameters in our catalog is \texttt{dflux\underline{ }per}, as discussed earlier in \S\ref{sec:parameters}. This quality flag has already been utilized to select reliable results. Stricter criteria, such as \texttt{dflux\underline{ }per}$<0.10$ or $<0.12$, can be applied to identify even more reliable sources, particularly for metal-poor stars. For the metal-poor catalog, we also include the extinction $A_{\mathrm{V}}$ used in this paper, as we should exercise caution with the low-latitude high extinction stars. Additionally, the catalog has been cross-matched with Gaia to obtain various parameters from Gaia DR3, such as proper motion and magnitude. The global and metal-poor catalogs, along with the code for correcting the systematic patterns in the spectra, are made publicly available.

\section{Summary}\label{sec:summary}
In this paper, we characterize the patterns of systematic errors present in the absolute-calibrated Gaia XP spectra, using a very large number of XP spectra and their best-fitting synthetic spectra based on model atmospheres. We find that those patterns depend on stellar colors, brightness, extinction, and metallicity. We present a simple NN that relates the systematic flux patterns with stellar color, magnitude, and extinction. The predicted patterns match those in the data very well. After correcting the wiggles, \texttt{FER\reflectbox{R}E} is applied to derive atmospheric parameters from corrected Gaia XP spectra. Our methodology is validated from the comparison with APOGEE DR17 parameters and stars with observations in the HST CALSPEC collection.

Compared to APOGEE, our estimation of atmospheric parameters is accurate in the temperature range $4000 \le T_{\mathrm{eff}} \le 7000$ K, with slight systematic errors and standard deviations around $-38 \pm 167$ K, $0.05 \pm 0.40$ dex, and $-0.12 \pm 0.19$ dex in $T_{\mathrm{eff}}$, $\log g$, and [M/H], respectively. The estimation of atmospheric parameters and spectra flux are both improved by correcting the systematic patterns. Our corrections improve the quality of the relative spectrophotometry of the Gaia XP data from $3.2\% - 3.7\%$ to $1.2\% - 2.4\%$, as verified against our models and the high-quality CALSPEC standards. Our results are also compared with other catalogs generated from XP data in the literature, and the metallicity is validated through the use of star cluster members. Finally, we publish our atmospheric parameters catalog of $68,394,431$ sources, with a metal-poor ($\mathrm{[M/H]} \le -2.5$) subset including $124,188$ stars. Our catalogs and flux-correction code are publicly available.

\section{Data availability}

Tables and codes are made publicly available at \url{https://doi.org/10.5281/zenodo.14028588}. The tables are also available in electronic form at the CDS via anonymous ftp to \url{cdsarc.u-strasbg.fr} (130.79.128.5) or via \url{http://cdsweb.u-strasbg.fr/cgi-bin/qcat?J/A+A/}.

\begin{acknowledgements}
    This study is supported by the National Key R\&D Program of China under grant Nos. 2023YFE0107800, 2024YFA1611900, and National Natural Science Foundation of China under grant Nos. 11988101, 12273055, 11927804. This study is also supported by International Partnership Program of Chinese Academy of Sciences Grant No.178GJZ2022040GC. \\
    
    XY and WW acknowledge the support from the China Scholarship Council. \\

    CAP acknowledges financial support from the Spanish Ministry MICIU projects PID2020-117493GB-I00 and PID2023-149982NB-I00. This research made use of computing time available on the high-performance computing systems at the Instituto de Astrofisica de Canarias. The authors are thankful for the technical expertise and assistance provided by the Spanish Supercomputing Network (Red Espanola de Supercomputacion), and the staff at the Instituto de Astrofisica de Canarias. \\

    CdB acknowledges support from a Beatriz Galindo senior fellowship (BG22/00166) from the Spanish Ministry of Science, Innovation and Universities. \\
    
    This work presents results from the European Space Agency (ESA) space mission Gaia. Gaia data are being processed by the Gaia Data Processing and Analysis Consortium (DPAC). Funding for the DPAC is provided by national institutions, in particular the institutions participating in the Gaia MultiLateral Agreement (MLA). The Gaia mission website is \url{https://www.cosmos.esa.int/gaia}. The Gaia archive website is \url{https://archives.esac.esa.int/gaia}. \\

    This job has made use of the Python package GaiaXPy, developed and maintained by members of the Gaia Data Processing and Analysis Consortium (DPAC), and in particular, Coordination Unit 5 (CU5), and the Data Processing Centre located at the Institute of Astronomy, Cambridge, UK (DPCI). \\
    
    Funding for the Sloan Digital Sky 
    Survey IV has been provided by the 
    Alfred P. Sloan Foundation, the U.S. 
    Department of Energy Office of 
    Science, and the Participating 
    Institutions. 
    
    SDSS-IV acknowledges support and 
    resources from the Center for High 
    Performance Computing  at the 
    University of Utah. The SDSS 
    website is www.sdss4.org.
    
    SDSS-IV is managed by the 
    Astrophysical Research Consortium 
    for the Participating Institutions 
    of the SDSS Collaboration including 
    the Brazilian Participation Group, 
    the Carnegie Institution for Science, 
    Carnegie Mellon University, Center for 
    Astrophysics | Harvard \& 
    Smithsonian, the Chilean Participation 
    Group, the French Participation Group, 
    Instituto de Astrof\'isica de 
    Canarias, The Johns Hopkins 
    University, Kavli Institute for the 
    Physics and Mathematics of the 
    Universe (IPMU) / University of 
    Tokyo, the Korean Participation Group, 
    Lawrence Berkeley National Laboratory, 
    Leibniz Institut f\"ur Astrophysik 
    Potsdam (AIP),  Max-Planck-Institut 
    f\"ur Astronomie (MPIA Heidelberg), 
    Max-Planck-Institut f\"ur 
    Astrophysik (MPA Garching), 
    Max-Planck-Institut f\"ur 
    Extraterrestrische Physik (MPE), 
    National Astronomical Observatories of 
    China, New Mexico State University, 
    New York University, University of 
    Notre Dame, Observat\'ario 
    Nacional / MCTI, The Ohio State 
    University, Pennsylvania State 
    University, Shanghai 
    Astronomical Observatory, United 
    Kingdom Participation Group, 
    Universidad Nacional Aut\'onoma 
    de M\'exico, University of Arizona, 
    University of Colorado Boulder, 
    University of Oxford, University of 
    Portsmouth, University of Utah, 
    University of Virginia, University 
    of Washington, University of 
    Wisconsin, Vanderbilt University, 
    and Yale University. \\

    This work made extensive use of TOPCAT \citep{2005ASPC..347...29T}.

\end{acknowledgements}

\bibliographystyle{aa}
\bibliography{bibliography}

\begin{appendix}

\onecolumn

\section{Comparisons with \cite{2023ApJS..267....8A} and \cite{2023MNRAS.524.1855Z}} \label{appendix_a}

\begin{figure*}[htbp!]
    \centering
    \resizebox{17cm}{17cm}
    {\includegraphics {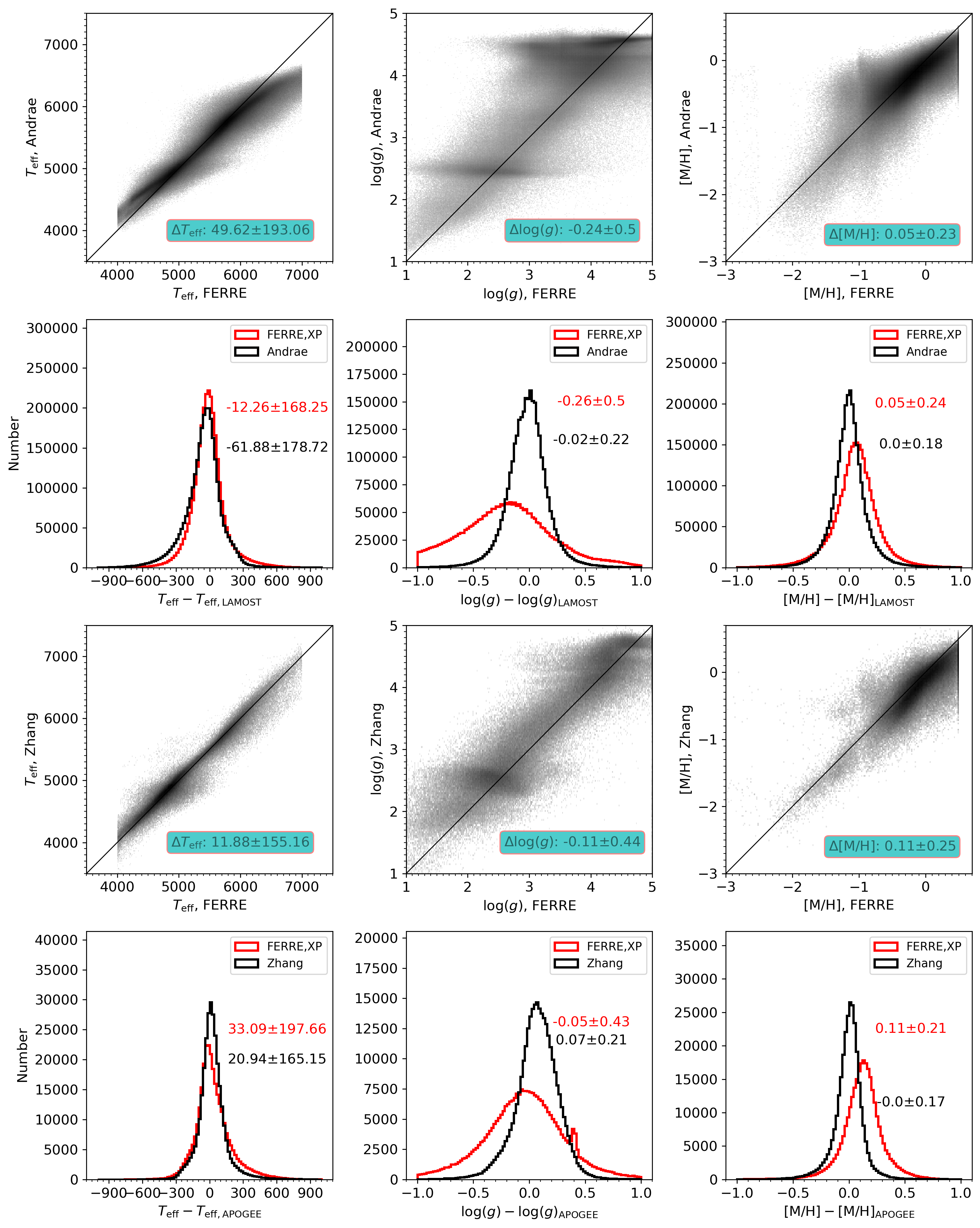}}
    \caption{Comparison of our catalog with other catalogs from literatures using LAMOST or APOGEE as references. The first row: comparison between our catalog and a catalog from \cite{2023ApJS..267....8A}, annotated with the means and standard deviations of the differences in $T_{\mathrm{eff}}$, $\log g$, and [M/H]. The second row: histograms in red showing the differences in atmospheric parameters between our catalog and the catalog from LAMOST, while histograms in black showing the differences between the catalog from \cite{2023ApJS..267....8A} and the one from LAMOST, with means and standard deviations annotated. The last two rows: similar to the first and second rows, we compare our catalog with the catalog from \cite{2023MNRAS.524.1855Z}, using APOGEE as a reference in those panels in the last row.}
    \label{fig:compare_catalogs}
\end{figure*}

The direct comparison between our results and those from \cite{2023ApJS..267....8A} and \cite{2023MNRAS.524.1855Z} are presented in the first and third rows of Figure \ref{fig:compare_catalogs}. The mean and standard deviation values of $\Delta T_{\mathrm{eff}}$, $\Delta \log g$, and $\Delta \mathrm{[M/H]}$ indicate that our results are more similar to those of \cite{2023MNRAS.524.1855Z} in $T_{\mathrm{eff}}$ and $\log g$, but align more closely with \cite{2023ApJS..267....8A} for [M/H]. Since machine learning results can exhibit artificially high performance on the training data, as the data is used to optimize model parameters, we use parameters from different surveys as reference points in our comparison. \cite{2023ApJS..267....8A} provides atmospheric parameters using \texttt{XGBoost} training on APOGEE. To make a fair comparison, we cross-match our catalog and theirs with LAMOST DR11 low-resolution catalog. With no additional cuts, there are almost $3$ million stars in common. We calculate $\Delta T_{\mathrm{eff}}, \Delta \log g, \Delta \mathrm{[M/H]}$ between our results and those from LAMOST, which are presented in the second row of Figure \ref{fig:compare_catalogs}. $\Delta T_{\mathrm{eff}}, \Delta \log g, \Delta \mathrm{[M/H]}$ between \cite{2023ApJS..267....8A} and LAMOST are also presented in those panels with different colors. We can see that we have a better estimation for $T_{\mathrm{eff}}$ when using the reference $T_{\mathrm{eff}}$ from LAMOST. However, our gravity estimation is significantly worse than theirs, and the metallicity from \cite{2023ApJS..267....8A} is also better than ours. In the last row of Figure \ref{fig:compare_catalogs}, we use APOGEE DR17 as a reference to calculate $\Delta T_{\mathrm{eff}}, \Delta \log g, \Delta \mathrm{[M/H]}$ between our catalog (or the catalog from \cite{2023MNRAS.524.1855Z}) with the catalog from APOGEE. From these histograms, using APOGEE as a reference, all three atmospheric parameters from \cite{2023MNRAS.524.1855Z} are better than ours. In the middle panel of the last row, a spike appears at $\log g - \log g_{\mathrm{APOGEE}} \sim 0.35 - 0.40$, which is caused by a problematic estimation of $\log g$ for low $T_{\mathrm{eff}}$ stars. A deeper analysis of the $T_{\mathrm{eff}}$ vs. $\log g - \log g_{\mathrm{APOGEE}}$ diagram reveals an overdensity at $\log g - \log g_{\mathrm{APOGEE}} \sim 0.35 - 0.40$ for low temperature stars ($T_{\mathrm{eff}} < 4500$).

\begin{figure*}[htbp!]
    \centering
    \resizebox{17cm}{17cm}
    {\includegraphics {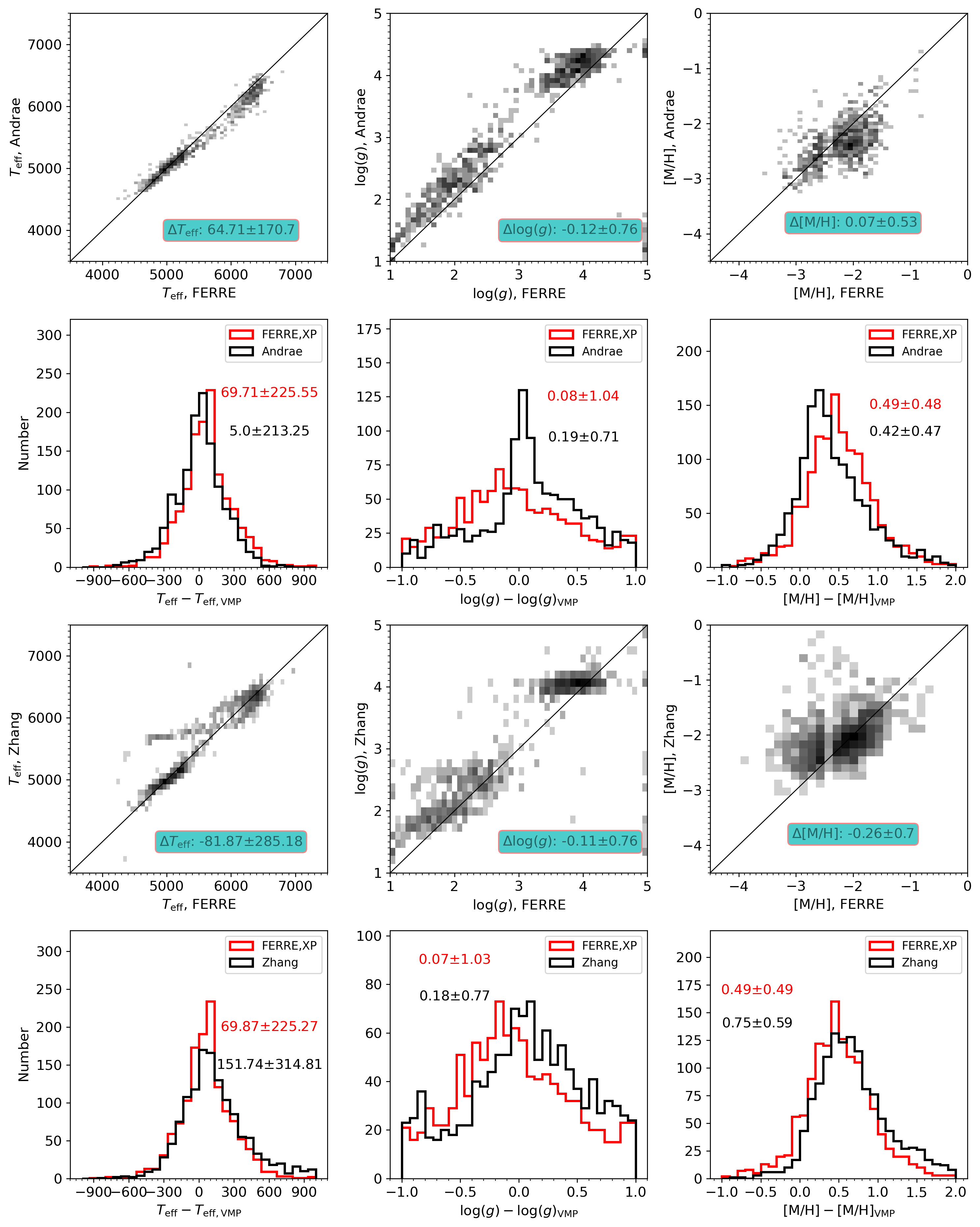}}
    \caption{Similar to Figure \ref{fig:compare_catalogs}, Comparison of our catalog with other catalogs from literatures using VMP stars as references.}
    \label{fig:compare_catalogs_vmp}
\end{figure*}

In addtion, we cross-match our catalog and the catalogs from the literature with VMP stars described in \S\ref{sec:metalpoor}. It should be pointed out that \cite{2023ApJS..267....8A} uses metal-poor stars to replenish their training sample, which contains many sources in common with our VMP selection. Therefore, comparison between our results and theirs for VMP stars is not fair. Not surprisingly, \cite{2023ApJS..267....8A} performs better in such comparison, but our method performs better compared to \cite{2023MNRAS.524.1855Z}, as shown in Figure \ref{fig:compare_catalogs_vmp}.

\section{Validating our catalog using star clusters} \label{appendix_b}

\begin{figure*}[htbp!]
    \centering
    \resizebox{17cm}{17cm}
    {\includegraphics {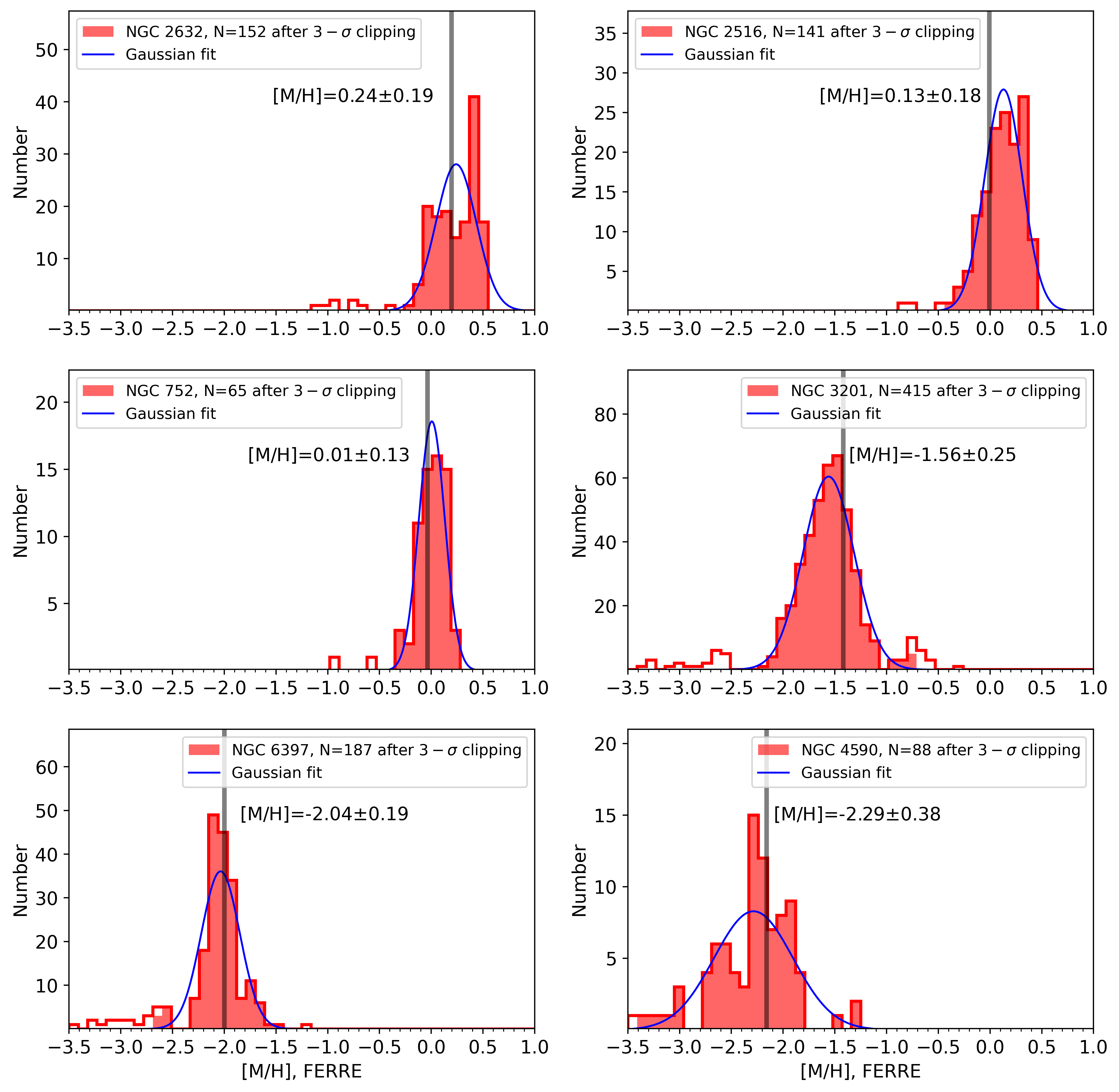}}
    \caption{The metallicity distribution from our catalog is shown for member stars of three open clusters (NGC 2632, NGC 2516, NGC 752) and three globular clusters (NGC 3201, NGC 6397, NGC 4590). In each panel, we display the metallicity distribution of member stars both before and after applying $3-\sigma$ clipping, depicted as step and filled histograms, respectively, with the number of remaining members annotated in the labels. Gaussian fits are overlaid as blue curves in all panels. The mean and standard deviation of [M/H] for each cluster are also presented in the respective panel. The black vertical lines indicate the metallicities of these star clusters as reported in the literature.}
    \label{fig:compare_scs}
\end{figure*}

The star clusters are selected primarily at random, ensuring a sufficient number of member candidates to construct the metallicity distribution, while also ensuring that each cluster has a single, well-defined metallicity. Member candidates are from \cite{2024A&A...686A..42H} and \cite{2021MNRAS.505.5978V} for OCs and GCs\footnote[8]{Table of member stars: \href{https://zenodo.org/records/4891252}{https://zenodo.org/records/4891252}}, respectively. To confirm the membership of stars, we use the membership probability $\texttt{Prob}>0.6$ and $\texttt{memberprob}=1$ for OCs and GCs, respectively. Similar to \cite{2024arXiv241019895H}, we select only giants (with our estimated $\log g < 3.5$) in the case of GCs, and fit the metallicity distribution for each cluster, using a Gaussian function after applying $3-\sigma$ clipping. The metallicity distributions for the three OCs and three GCs are shown in Figure \ref{fig:compare_scs}. In this figure, we include all the candidates, but highlight the distribution after applying $3-\sigma$ clipping. The number of member stars for each star cluster (after $3-\sigma$ clipping) and Gaussian fit to the clipped distribution are shown in each panel as well.

From each panel in Figure \ref{fig:compare_scs}, stars within the star cluster present strong compatibility in [M/H]. The metallicity of OCs is close to $0$ but still shows variation among different OCs, while the metallicity of GCs falls in the range $-2.3 \sim -1.6$. We also compare the metallicity distribution with values from literature in Figure \ref{fig:compare_scs}, indicated by the black vertical lines. For OCs, metallicities are adopted from \cite{2021MNRAS.504..356D}, while for GCs, we use the values from \cite{1998AJ....116..765G} for NGC 3201, \cite{2000A&A...361...92C} for NGC 6397, and \cite{2005AJ....129..251L} for NGC 4590.

\end{appendix}

\end{document}